\documentclass[a4paper,10pt]{article}
\usepackage{amsmath,amssymb,amsthm,amsfonts,epsf}
\usepackage{graphicx,eepic}
\usepackage{latexsym}

\def\b{\begin{eqnarray}}
\def\e{\end{eqnarray}}
\def\n{\noindent}
\def\openone{\leavevmode\hbox{\small1\kern-0.355em\normalsize1}}
\def\biglb{\big[\hspace*{-.7mm}\big[}
\def\bigrb{\big]\hspace*{-.7mm}\big]}

\def\Res{\mathop{\mbox{Res}\,}\limits}

\def\bigglb{\bigg[\hspace*{-.7mm}\bigg[}
\def\biggrb{\bigg]\hspace*{-.7mm}\bigg]}
\def\Bigglb{\Bigg[\hspace*{-1.4mm}\Bigg[}
\def\Biggrb{\Bigg]\hspace*{-1.3mm}\Bigg]}
\def\bbbc{{\Bbb C}}

\def\bbbr{{\Bbb R}}

\def\im{\mbox{Im}\,}

\def\bigrbt{\mathop{\bigrb }\limits_{\widetilde{\;}}}

\newtheorem{lemma}{Lemma}
\newtheorem{cor}{Corrollary}
\newtheorem{remark}{Remark}
\newtheorem{example}{Example}
\newtheorem{proposition}{Proposition}
\allowdisplaybreaks
\begin{document}

\begin{center}

{\LARGE\textbf{Generalised Fourier transform for the Camassa-Holm
hierarchy
\\}} \vspace {10mm} \vspace{1mm} \noindent

{\large \bf Adrian Constantin$^{a,\dag}$}, {\large \bf  Vladimir
S. Gerdjikov$^{b,\ddag}$} and \\ {\large \bf Rossen I.
Ivanov$^{a,\ast,}$}\footnote{On leave from the Institute for
Nuclear Research and Nuclear Energy, Bulgarian Academy of
Sciences, Sofia, Bulgaria.} \vskip1cm \n \hskip-.3cm
\begin{tabular}{c}
\hskip-1cm $\phantom{R^R} ^{a}${\it School of Mathematics, Trinity
College Dublin,}
\\ {\it Dublin 2, Ireland} \\
$\phantom{R^R}^{b}${\it Institute for Nuclear Research and Nuclear
Energy,}\\ {\it Bulgarian Academy of Sciences,} \\
{\it 72 Tzarigradsko chaussee, 1784 Sofia,
Bulgaria} \\
\\{\it $^\dag$e-mail: adrian@maths.tcd.ie}
\\{\it $^\ddag$e-mail: gerjikov@inrne.bas.bg}
\\ {\it $^\ast$e-mail: ivanovr@maths.tcd.ie}
\\
\hskip-.8cm
\end{tabular}
\vskip1cm
\end{center}

\vskip1cm

\begin{abstract}
\n The squared eigenfunctions of the spectral problem associated
to the Camassa-Holm equation represent a complete basis of
functions, which helps to describe the Inverse Scattering
Transform for the Camassa-Holm hierarchy as a Generalised Fourier
transform. The main result of this work is the derivation of the
completeness relation for the squared solutions of the
Camassa-Holm spectral problem. We show that all the fundamental
properties of the Camassa-Holm equation such as the integrals of
motion, the description of the equations of the whole hierarchy
and their Hamiltonian structures can be naturally expressed making
use of the completeness relation and the recursion operator, whose
eigenfunctions are the squared solutions.



\end{abstract}

\newpage

\section{Introduction}\label{sec:1}

In this introductory section we shall give a brief account of the
basic results about the spectral problem related to the Camassa-Holm equation
(CH) \cite{CH93}. The CH
equation
\begin{equation}\label{eq1}
 u_{t}-u_{xxt}+2\omega u_{x}+3uu_{x}-2u_{x}u_{xx}-uu_{xxx}=0,
\end{equation}
where $\omega$ is a real constant, arises as a compatibility condition of
two linear problems \cite{CH93}

\b \label{eq3} \Psi_{xx}&=&\Big(\frac{1}{4}+\lambda
(m+\omega)\Big)\Psi
 \\\label{eq4}
\Psi_{t}&=&\Big(\frac{1}{2\lambda}-u\Big)\Psi_{x}+\frac{u_{x}}{2}\Psi+\gamma\Psi
\e where $m\equiv u-u_{xx}$ and $\gamma$ is an arbitrary constant.
We will use the freedom provided by the presence of $\gamma$ for a proper normalization of the
eigenfunctions.

The CH equation models just like the Korteweg-de Vries (KdV)
equation the propagation of two-dimensional shallow water waves
over a flat bed. It also arises in the study of the propagation of
axially symmetric waves in hyperelastic rods \cite{D98, CS00-2}
and its high-frequency limit models nematic liquid crystals
\cite{HS91, BC05}; more about the physical applications of this
equation can be found e.g. in \cite{CH93, J02, J03,DGH03, DGH04,
GH03, I07, SS07}. Both KdV and CH are integrable systems
\cite{ZMNP, BBS98, C98, CM99, C01, L02} (see also \cite{CH93,
FF81, FOR96}), but while all smooth data yield solutions of the
KdV equation existing for all times, certain smooth initial data
for CH lead to global solutions and others to breaking waves: the
solution remains bounded but its slope becomes unbounded in finite
time (see \cite{CE98, C00, BC07}).  The solitary waves of KdV are
smooth solitons, while the solitary waves of CH, which are also
solitons, are smooth if $\omega > 0$ \cite{CH93,J03,PI,PII,PIII}
and peaked (called ``peakons" and representing weak solutions) if
$\omega=0$ \cite{CH93,CE98-2,BBS98,BBS99,CM00,L05-2}. Both
solitary wave forms for CH are stable \cite{CS00, CM01, CS02}. The
CH equation arises also as a geodesic equation on the
diffeomorphism group (if $\omega=0$) \cite{C00, CK03, CK06, K04}
and on the Bott-Virasoro group (if $\omega > 0$) \cite{M98,
CKKT07}.

The inverse scattering transform (IST) for the CH equation and the
related Riemann-Hilbert problem are considered e.g. in
\cite{C01,CGI,K05,dMonv}.

The IST can be realized as a generalized Fourier transform.The
complete basis of functions is represented by the squares of the
fundamental (Jost) solutions of the corresponding spectral
problem. For the famous Zakharov-Shabat spectral problem (and its
various generalizations) the problem is studied in detail and the
results can be found in several important works, such as
\cite{K76,E81,G86,GI92,GY94}. The generalized Fourier transform for the Sturm-Liouville spectral
problem, associated to the fundamentally important KdV equation is
also well studied and in this relation we can mention the books
and articles \cite{B74,Cal,K80,E81,IKK94,K98,K03}.

The CH spectral problem (\ref{eq3}) is of
a weighted Sturm-Liouville-type. Our aim will be to construct the generalized
Fourier transform which linearizes the CH equation as well as all
equations of the whole CH hierarchy of integrable equations.
Our main result is the derivation of the completeness relation for
the squared solutions of (\ref{eq3}). We show that all the
fundamental properties of the CH equation such as the integrals of
motion, the description of the higher CH-type equations and their
Hamiltonian structures can be naturally expressed making use of
the recursion operator $L_\pm$ and the completeness relation.
In fact, the squared solutions are eigenfunctions of $L_\pm$ and
the completeness of the squared solutions is the spectral
decomposition of $L_\pm$.

In Section  \ref{sec:2} are given all the necessary mathematical
preliminaries about the spectral problem (\ref{eq3}). This
includes the definition of the Jost solutions and two sets of
scattering data, as well as the time evolution of the scattering
data.

In Section \ref{sec:3} are presented the asymptotics of the Jost
solutions for large values of the spectral parameter ($|k|\to
\infty$). The main difference with respect to the standard
spectral problem, given by the Schr\"odinger equation, (e.g. like
in the KdV case) lies in the fact that (\ref{eq3}) is a weighted
spectral problem, which requires different asymptotic expansions.

The completeness relations of the eigenfunctions and the squared
eigenfunctions of the spectral problem (\ref{eq3}) are presented
in Section  \ref{sec:4}. The last one gives the possibility of
expansion of an arbitrary function of the specified class over the
complete basis.

The Wronskian relations, derived in Section \ref{sec:5} allow to
compute the generalised Fourier coefficients for the potential of
our spectral problem and its variation.

The symplectic variables, i.e. the variables, expressed in terms
of the squared solutions and satisfying the cannonical relations,
with respect to a certain skew-product are given in Section
\ref{sec:6}.

The recursion operator (whose eigenfunctions are the squared
solutions) is computed in Section  \ref{sec:7}.

In Section \ref{sec:8} the whole Camassa-Holm hierarchy is
constructed. The time-evolution of the scattering data for the
hierarchy is obtained.

In Section \ref{sec:9} the Hamiltonian structure of the
Camassa-Holm hierarchy is explored. The canonical Hamiltonians are
expressed both via the potential of the spectral problem and the
scattering data. The hierarchy of Poisson structures and
action-angle variables with respect to these structures is also
obtained in this section.

Finally, in Section  \ref{sec:10} the Inverse Scattering Transform
for the CH hierarchy is outlined in the framework presented
earlier for the CH equation alone \cite{CGI}.

\section{Preliminaries}\label{sec:2}

In general, there exists an infinite sequence of conservation laws
(multi-Hamiltonian structure) $H_n[m]$, $n=0,\pm1, \pm2,\ldots$,
\cite{CH93,FS99,R02,I05,FF81,L05} such that \b\label{H1a}
H_1&=&\frac{1}{2}\int mu \text{d}x, \\\label{H2a}
H_2&=&\frac{1}{2}\int (u^3+uu_x^2+2\omega u^2)
\text{d}x, \\
(\partial-\partial^{3})\frac{\delta H_{n}[m]}{\delta m}&=&(2\omega
\partial +m\partial+\partial m)\frac{\delta H_{n-1}[m]}{\delta m}.
\label{eq2a}\e

The CH equation can be written as
\begin{equation}\label{eq1a}
 m_{t}=\{m, H_{1}\},
\end{equation}
where the Poisson bracket is defined as
\begin{equation}\label{PB}
\{A,B\}\equiv -\int \frac{\delta A}{\delta m}(2\omega
\partial+ m\partial+\partial m)\frac{\delta B}{\delta m}\text{d}x,
\end{equation}
or in more obvious antisymmetric form.
\begin{equation}\label{PBa}
\{A,B\}=-\int (\omega+m)\Big(\frac{\delta A}{\delta m}\partial
\frac{\delta B}{\delta m}- \frac{\delta B}{\delta m}\partial
\frac{\delta A}{\delta
 m}\Big)\text{d}x.
\end{equation}

For simplicity we will consider the case where $m$ is a Schwartz
class function, $\omega >0$ and $m(x,0)+\omega > 0$. Then
$m(x,t)+\omega > 0$ for all $t$ \cite{C01}.  Let
$k^{2}=-\frac{1}{4}-\lambda \omega$, i.e. \b \label{lambda}
\lambda(k)= -\frac{1}{\omega}\Big( k^{2}+\frac{1}{4}\Big).\e

The spectrum of the problem (\ref{eq3}) under these conditions is
described in \cite{C01}. The continuous spectrum in terms of $k$
corresponds to $k$ -- real. The discrete spectrum consists of
finitely many points $k_{n}=i\kappa _{n}$, $n=1,\ldots,N$ where
$\kappa_{n}$ is real and $0<\kappa_{n}<1/2$.

A basis in the space of solutions of (\ref{eq3}) can be introduced
by the analogs of the Jost solutions of eq. (\ref{eq1}),
$f^+(x,k)$ and $\bar{f}^+(x,\bar{k})$. For all real $k\neq 0$ it
is fixed by its asymptotic when $x\rightarrow\infty$ \cite{C01},
see also \cite{ZMNP}: \b \label{eq6} \lim_{x\to\infty }e^{-ikx}
f^+(x,k)= 1, \e

\n   Another basis can be introduced, $f^-(x,k)$
and$\bar{f}^-(x,\bar{k})$ fixed by its asymptotic when
$x\rightarrow -\infty$ for all real $k\neq 0$: \b \label{eq6'}
\lim_{x\to -\infty }e^{ikx} f^-(x,k)= 1, \e

\n Since $m(x)$ and $\omega$ are real one gets that if $f^+(x,k)$
and $f^-(x,k)$ are solutions of (\ref{eq1}) then
\begin{equation}\label{eq:inv}
 \bar{f}^+(x,\bar{k}) = f^+(x,-k), \qquad \mbox{and} \qquad
 \bar{f}^-(x,\bar{k}) = f^-(x,-k),
\end{equation}
are also solutions of (\ref{eq1}). The relations (\ref{eq:inv})
are known as involutions.

\n In particular, for real $k\neq 0$ we get:

\b \label{eq5aa} \bar{f}^{\pm}(x,k)=f^{\pm}(x, -k), \e

\n and the vectors of the two bases are related
\footnote{According to the notations used in \cite{CGI,CI06}
$f^+(x,k)\equiv \bar{\psi}(x,\bar{k})$, $f^-(x,k)\equiv
\varphi(x,k)$.}:

\b \label{eq8} f^{-}(x,k)=a(k)f^+(x,-k)+b(k)f^+(x,k), \qquad
\mathrm{Im}\phantom{*} k=0. \e

\n The Wronskian $W(f_{1},f_{2})\equiv
f_{1}\partial_{x}f_{2}-f_{2}\partial_{x}f_{1}$ of any pair of
solutions of (\ref{eq3}) does not depend on $x$. Therefore

\b \label{eq9} W(f^-(x,k), f^-(x,-k))= W(f^+(x,-k), f^+(x,k))=2ik
\e

\n  Computing the Wronskians $W(f^-,f^+)$ and $W(\bar{f}^+,f^-)$
and using (\ref{eq8}), (\ref{eq9}) we obtain: \b
\label{eq10a}a(k)&=&(2ik)^{-1} W(f^-(x,k), f^+(x,k)), \\
\label{eq10b} b(k)&=&-(2ik)^{-1} W(f^-(x,k), f^+(x,-k)).\e

\n From (\ref{eq8}) and (\ref{eq9}) it follows that for real $k$

\b \label{eq10a1} a(k)a(-k)-b(k)b(-k)=1. \e

It is well known \cite{C01} that $f^+(x,k)e^{-ikx}$ and
$f^-(x,k)e^{ikx}$ have analytic extensions in the upper half of
the complex $k$-plane. Likewise
$\bar{f}^+(x,\bar{k})e^{i\bar{k}x}$ and $\bar{f}^-(x,\bar{k})
e^{-i\bar{k}x}$ allow analytic extension in the lower half of the
complex $k$-plane. An important consequence of these properties is
that $a(k)$ also allows analytic extension in the upper half of
the complex $k$-plane and
\begin{equation}\label{eq:inv1}
\bar{a}(\bar{k}) = a(-k), \qquad \bar{b}(\bar{k}) = b(-k),
\end{equation}
As a result  (\ref{eq10a1}) acquires the form: \b \label{eq10}
|a(k)|^{2}-|b(k)|^{2}=1. \e

\n In other words the relation (\ref{eq10a}) is valid in the upper
half plane $\mathrm{Im}\phantom{*} k\geq 0$ \cite{CI06,CGI}, while
(\ref{eq10b}) makes sense only on the real line
$\mathrm{Im}\phantom{*} k = 0$. In analogy with the spectral
problem for the KdV equation, one can introduce the quantities
$\mathcal{T}(k)=a^{-1}(k)$ (transmission coefficient) and
$\mathcal{R}^{\pm}(k)=b(\pm k)/a(k)$, (reflection coefficients --
to the right with superindex ($+$) and to the left with superindex
($-$) respectively).  From (\ref{eq10}) it follows that

\b \label{eq13} |\mathcal{T}(k)|^{2}+|\mathcal{R}^{\pm}(k)|^{2}=1.
\e

\n It is sufficient to know $\mathcal{R}^{\pm}(k)$ only on the
half line $k>0$, since from (\ref{eq:inv1}), $\bar{a}(k)=a(-k)$,
$\bar{b}(k)=b(-k)$ and thus
$\mathcal{R}^{\pm}(-k)=\bar{\mathcal{R}}^{\pm}(k)$. Also, from
(\ref{eq13})

\b \label{eq13a} |a(k)|^{2}=(1-|\mathcal{R}^{\pm}(k)|^{2})^{-1},
\e

\n One can show that $\mathcal{R}^{\pm}(k)$ uniquely determines
$a(k)$ \cite{CI06}.

At the points $\kappa_n$ of the discrete spectrum, $a(k)$ has
simple zeroes \cite{C01}, i.e.:
\begin{equation}\label{eq:a-n}
    a(k) = (k-i\kappa_n)\dot{a}_n +\frac{1}{2} (k-i\kappa_n)^2\ddot{a}_n
    + \cdots,
\end{equation}
and  the Wronskian $W(f^-,f^+)$, (\ref{eq10a}) vanishes. Thus
$f^-$ and $f^+$ are linearly dependent:

\b \label{eq200} f^-(x,i\kappa_n)=b_nf^+(x,i\kappa_n).\e

\n  In other words, the discrete spectrum is simple, there is only
one (real) linearly independent eigenfunction, corresponding to
each eigenvalue $i\kappa_n$, say

\b \label{eq201}f_n^-(x)\equiv f^-(x,i\kappa_n)\e

\n From (\ref{eq201}) and (\ref{eq6}), (\ref{eq6'}) it follows
that $f_n^-(x)$ falls off exponentially for $x\to\pm\infty$, which
allows one to show that $f_n(x)$ is square integrable. Moreover,
for compactly supported potentials $m(x)$ (cf. (\ref{eq200}) and
(\ref{eq8}))

\b \label{eq202} b_n= b(i\kappa_n), \qquad
b(-i\kappa_n)=-\frac{1}{b_n} .\e

\n One can argue \cite{ZMNP}, that the results from this case can
be extended to Schwarz-class potentials by an appropriate limiting
procedure.

\n The asymptotic of $f_n^-$, according to (\ref{eq5aa}),
(\ref{eq6}), (\ref{eq200}) is

\b \label{eq203} f_n^-(x)&=&e^{\kappa_n x}+o(e^{\kappa_n x}),
\qquad x\rightarrow -\infty;
 \\\label{eq204}
f_n^-(x)&=& b_n e^{-\kappa_n x}+o(e^{-\kappa_n x}), \qquad
x\rightarrow \infty. \e

\n The sign of $b_n$ obviously depends on the number of the zeroes
of $f_n^-$. Suppose that
$0<\kappa_{1}<\kappa_{2}<\ldots<\kappa_{N}<1/2$. Then from the
oscillation theorem for the Sturm-Liouville problem \cite{B},
$f_n^-$ has exactly $n-1$ zeroes. Therefore

\b \label{eq205} b_n= (-1)^{n-1}|b_n|.\e

The sets

\b \label{eq206} \mathcal{S^{\pm}}\equiv\{
\mathcal{R}^{\pm}(k)\quad (k>0),\quad \kappa_n,\quad
C_n^{\pm}\equiv\frac{b_n^{\pm1}}{i\dot{a}_n},\quad n=1,\ldots N\}
\e

\n are called scattering data. Throughout
this work the dot stands for a derivative
with respect to $k$ and $\dot{a}_n\equiv \dot{a}(i\kappa_n)$,
$\ddot{a}_n\equiv \ddot{a}(i\kappa_n)$, etc. In what follows we
will show that each set -- $\mathcal{S^{+}}$ or $\mathcal{S^{-}}$
of scattering data uniquely determines the potential $m(x)$. The
derivation is similar to those for other integrable systems, e.g.
\cite{ZMNP,E81,IKK94,ZF71}.

The time evolution of the scattering data can be easily obtained
as follows. From (\ref{eq8}) with $x\rightarrow\infty$ one has

\b \label{eq14} \lim_{x\to\infty}\left( f^-(x,k)- a(k)e^{-ikx}-
b(k)e^{ikx}\right) =0. \e The substitution of $f^-(x,k)$ into
(\ref{eq4}) with $x\rightarrow\infty$ gives

\b \label{eq15}
\lim_{x\to\infty}(f^-_{t}-\frac{1}{2\lambda}f^-_{x}+\gamma f^-)=0.
\e

\n From (\ref{eq14}), (\ref{eq15}) with the choice
$\gamma=\gamma_{-}=ik/2\lambda$ we obtain

\b \label{eq16} a_t(k,t)&=&0,
 \\\label{eq17}
b_t(k,t)&=& \frac{i k}{\lambda }b(k,t). \e

\n Thus
 \b \label{eq18} a(k,t)=a(k,0), \qquad b(k,t)=b(k,0)e^{\frac{i
k}{\lambda }t}; \e

\b \label{eq19} \mathcal{T}(k,t)=\mathcal{T}(k,0), \qquad
\mathcal{R}^{\pm}(k,t)=\mathcal{R}^{\pm}(k,0)e^{\pm\frac{i
k}{\lambda }t}. \e

Similarly, one can substitute $f^+(x,k)$ as $x\rightarrow-\infty$
\b \label{eq14a} \lim_{x\to\infty}\left( f^+(x,k)-a(k)e^{ikx}+
b(-k)e^{-ikx}\right) =0 \e into (\ref{eq4}). Then the choice of
the constant is $\gamma=\gamma_{+}=-ik/2\lambda$ and the final
result (\ref{eq18}) is, of course the same.

In other words, $a(k)$ is independent on $t$ and will serve as a
generating function of the conservation laws.

The time evolution of the data on the discrete spectrum is found
as follows. $i\kappa_n$ are zeroes of $a(k)$, which does not
depend on $t$, and therefore $\kappa_{n,t}=0$. From (\ref{eq202})
and (\ref{eq17}) one can find
 \b \label{eq207} C_n^{\pm} (t)= C_n^{\pm} (0)\exp \Big(\pm\frac{4\omega
\kappa_n}{1-4\kappa_n^2}t\Big). \e

\section{Asymptotics of the Jost solutions for $|k|\to \infty$}\label{sec:3}

The analyticity properties of the Jost solutions and of $a(k)$
play an important role in our considerations. We will need also
the asymptotics of the Jost solutions for $|k|\rightarrow \infty$
which have the form \cite{CGI,CI06}

\b\label{eq21aa} f^{+} (x,k)=e^{ i k
x+ik\int_{\infty}^x(\sqrt{q(y)/\omega} -1)dy}
\Big[\Big(\frac{\omega}{q(x)}\Big)^{1/4}+O\Big(\frac{1}{k}\Big)\Big],\e

\b\label{eq22} f^-(x,k)=e^{-ikx-ik\int_{-\infty}^x(\sqrt{q(y)/
\omega}-1)dy}
\Big[\Big(\frac{\omega}{q(x)}\Big)^{1/4}+O\Big(\frac{1}{k}\Big)\Big],\e

\n where, for simplicity $q(x)\equiv m(x)+\omega$.

An immediate consequence of the above formulae and (\ref{eq10a})
is:
\begin{equation}\label{eq:a-as}
    \lim_{k\to \infty} a(k) e^{i\alpha k} = 1, \qquad k\in \bbbc_+,
\end{equation}
where
\begin{equation}\label{eqi8}
\alpha= \int _{-\infty}^{\infty}\Big(\sqrt{\frac{q(x)}
{\omega}}-1\Big)\text{d}x.
\end{equation}
Since $a(k)$ is $t$-independent, then $\alpha$, as well as all the
coefficients $I_k$ in the asymptotic expansion
\begin{equation}\label{eqi7}
 \ln a(k)= -i\alpha k+\sum_{s=1}^{\infty}\frac{I_{s}}{k^{2s+1}},
\end{equation}
must be integrals of motion. The integral $\alpha$ is the unique Casimir function for the CH,
\cite{KM03}. One can easily check that $\{m,\alpha\}\equiv 0$, for
the uniqueness argument and for the geometric interpretation see
\cite{KM03}.

The densities, $p_{s}$ of  $I_{s}=\int _{-\infty}^{\infty}p_{s}
\text{d}x$ can be expressed in terms of $m(x)$ using a set of
recurrent relations  obtained  in \cite{R02,I05,CI06}.

Using the analyticity properties of $a(k)$ one can prove that it
satisfies the following dispersion relation ($k\in \bbbc_+$):
\begin{equation}\label{eqi21}
\ln a(k)=-i\alpha k +\sum _{n=1}^{N}
\ln\frac{k-i\kappa_n}{k+i\kappa_n} - \frac{k}{\pi i} \int
_{0}^{\infty} \frac{\ln (1-|\mathcal{R}^\pm(k')|^2)}
{{k'}^2-k^2}\text{d}k'.
\end{equation}
where $i\kappa_n$ are the zeroes of $a(k)$. Thus we are able to
recover the function $a(k)$ in its whole domain of analyticity
$\bbbc_+$ knowing just its modulus $|a(k)|$ or the reflection
coefficient $\mathcal{R}^{+}(k)$, for real $k$, and the location
of its zeroes, see (\ref{eq13a}).

The dispersion relation (\ref{eqi21}) allows one to express the
integrals of motion also in terms of the scattering data
\cite{CI06}:
\begin{equation}\label{eq:I-n}
I_s =  \frac{1}{\pi i} \int_0^\infty \ln (1-|\mathcal{R}^\pm (k)
|^2)\, k^{2s} \text{d}k - \sum_{n=1}^N \frac{2i(-1)^s
\kappa_n^{2s+1}}{2s+1}.
\end{equation}
These are known as the trace identities. In addition the integral
$\alpha$ is expressed through the scattering data as follows. Note
that for $k=i/2$, $\lambda(i/2)=0$ from (\ref{lambda}). In this
case therefore the spectral problem (\ref{eq3}) does not depend on
$m$, and the eigenfunctions are equal to their asymptotics:
$f^{\pm}(x, i/2)=e^{\mp \frac{x}{2}}$. From (\ref{eq10a}) we
obtain $a(i/2)=1$ and from (\ref{eqi21}) for $k=i/2$ we have
\begin{equation}\label{eqi24}
\alpha=\sum_{n=1}^{N}\ln\Big(\frac{1+2\kappa_n}{1-2\kappa_n}
\Big)^2+ \frac{4}{\pi }\int _{0}^{\infty}\frac{\ln ( 1 -
|\mathcal{R}^{+}( \widetilde{k})|^2)}
 {4\widetilde{k}^2+1}\text{d}\widetilde{k}.
\end{equation}

\section{Completeness relations for the Jost solutions}\label{sec:4}

Here we outline the spectral properties of the linear problem
(\ref{eq3}).

Let us consider the function

\b \label{eq:res1} R_1(x,y,k)\equiv
\frac{f^-(x,k)f^+(y,k)}{a(k)}\theta(y-x)+
\frac{f^-(y,k)f^+(x,k)}{a(k)}\theta(x-y)\label{eq221 }\e where
$\theta (x)$ is the step function.

\begin{lemma}\label{lem-1}
\begin{description}
\item [i)] $R_1(x,y,k)$ is an analytic function of $k$ for $k\in
\bbbc_+$; \item [ii)] $R_1(x,y,k)$ has simple poles for
$k=i\kappa_n$; \item [iii)] $R_1(x,y,k)$ is a kernel of a bounded
integral operator for $\im k>0$. For $\im k=0$, $R_1(x,y,k)$ is a
kernel of a unbounded integral operator. \item[iv)] $R_1(x,y,k)$
satisfies the equation:
\begin{equation}\label{eq:res}
    \frac{\text{d}^2R_1}{\text{d}x^2} - \left( \frac{1}{4}
    +\lambda q(x) \right) R_1(x,y,k) =4ik\delta(x-y).
\end{equation}
\end{description}
\end{lemma}

\begin{proof} i) and ii) are obvious.

iii) For $\im k>0$ the statement follows from the definitions
(\ref{eq6}) and (\ref{eq6'}) of the Jost solutions, which ensure
that $R_1(x,y,k)$ falls off exponentially for all $x,y\to\pm
\infty$. The same arguments for $\im k=0$ can only ensure that
$R_1(x,y,k)$ is a bounded function for $x,y \to \pm \infty$.

iv) is a consequence of (\ref{eq3}), (\ref{eq10a}) and  the fact
that \[ d\theta(x-y)/dx = \delta(x-y).\]

\end{proof}

\begin{remark} \label{rem-res}
The function $R_1(x,y,k)/(4ik)$ is a kernel of the resolvent for
the linear problem (\ref{eq3}).

\end{remark}

The explicit expression for the resolvent (\ref{eq:res1})  can be
used to prove the following

\begin{proposition}\label{pr1}
The Jost solutions of (\ref{eq3}) satisfy the completeness
relation: \b\label{eq220} \frac{\omega}{q(x)}
\delta(x-y)=\frac{1}{2\pi}\int_{-\infty}^{\infty}
\frac{f^-(x,k)f^+(y,k)}{a(k)}{\text d}k +
\sum_{n=1}^{N}C_n^{\pm}f_n^{\pm}(x)f_n^{\pm}(y) \e By
$\int_{-\infty}^{\infty}dk$ is meant
$\lim_{L\rightarrow\infty}\int_{-L}^{L}\text{d}k$ and
$f_n^{\pm}(x)\equiv f^{\pm}(x,i\kappa_n)$.
\end{proposition}

\begin{proof}
Consider  the contour integral
\begin{equation}\label{eq:J1}
\mathcal{J}_1(x,y) = \frac{1}{2\pi i} \oint_{\gamma_+} R_1(x,y,k).
\end{equation}
where the contour $\gamma_+$ is shown on Fig. \ref{fig:1}.

\begin{figure}[t]
\centering \setlength{\unitlength}{0.2100pt}
\ifx\plotpoint\undefined\newsavebox{\plotpoint}\fi
\sbox{\plotpoint}{\rule[-0.175pt]{0.350pt}{0.350pt}}%
\special{em:linewidth 0.3pt}%
\begin{picture}(1440,1440)(0,0)
\put(20,720){\line(1,0){1420}} \put(1410,720){\vector(1,0){20}}
\put(720,10){\line(0,1){1430}} \put(720,1430){\vector(0,1){20}}
\put(1020,750){\vector(1,0){20}} \put(1020,690){\vector(1,0){20}}
\put(720,690){\arc{1320}{0}{3.14159}} \put(1380,1380){\circle{75}}
\put(1365,1365){$k$} \put(1186.7,1216.7){\vector(-1,1){20}}
\put(1186.7,223.3){\vector(-1,-1){20}}
\put(1186.7,183.3){$\gamma_{-,\infty}$}
\put(1186.7,1236.7){$\gamma_{+,\infty}$}
\put(60,690){\line(1,0){1320}}
\put(720,750){\arc{1320}{-3.14159}{0}}
\put(60,750){\line(1,0){1320}}
\end{picture}
\caption{The contours $\gamma_\pm = \bbbr \cup \gamma_{\pm,
\infty}$ of integrations.\protect\label{fig:1}}
\end{figure}
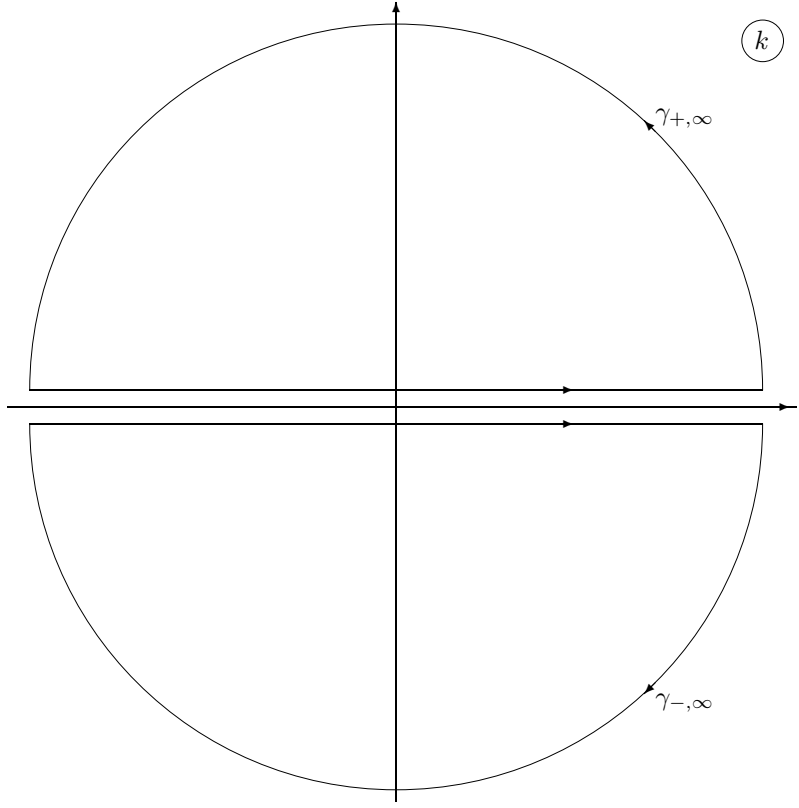

\n From the Cauchy residue theorem it follows that:
\begin{eqnarray}\label{eq:J-res}
\mathcal{J}_1(x,y) = \sum_{n=1}^N \Res_{k=i\kappa_n} R_1(x,y,k)
 =i \sum_{n=1}^N
C_n^{\pm}f_n^{\pm}(x)f_n^{\pm}(y).
\end{eqnarray}
In the evaluation of the residues we made use of (\ref{eq200}) and
of the fact that $\theta(x-y) + \theta(y-x)=1$.

Next we evaluate the integral $\mathcal{J}_1(x,y)$ by integrating
along the contour. For the integration along the infinite
semicircle we need the asymptotic of $R_1(x,y,k)$ for
$|k|\rightarrow \infty$. From  (\ref{eq21aa}) and (\ref{eq22}) we
get \b\label{eq260} R_1(x,y,k)=e^{ik\int_x^y
\sqrt{\frac{q(s)}{\omega}}d s}\Big[\Big(\frac{\omega}{q(x)}
\Big)^\frac{1}{4} \Big(\frac{\omega}{q(y)}\Big)^\frac{1}{4}
+o(1/k)\Big]\theta(y-x)+(x\leftrightarrow y)\e

Only the leading terms in (\ref{eq260}), which are entire
functions of $k$ contribute to the integral. This allows us to
deform the infinite semicircle until it coincides with the real
$k$-axis. Then the integration over $k$ is easily performed with
the result:
\begin{equation}\label{eq:J8}
\mathcal{J}_{1,\infty} (x,y) = \frac{i\omega}{q(x)} \delta(x-y).
\end{equation}

To evaluate the integral over the real axis
$\mathcal{J}_{1,R}(x,y)$ we will use the the fact that
$R_1(x,y,k)$ can be written in the form \b R_1(x,y,k)&=&
\frac{f^-(x,k)f^+(y,k)}{a(k)} + N_1(x,y,k)\theta(x-y),
\label{eq270-1}\e where \b N_1(x,y,k)\equiv
\frac{1}{a(k)}[f^-(x,k)f^+(y,k)-f^+(x,k)f^-(y,k)]\label{eq280}\e
is an odd function of $k$ and does not contribute to
$\mathcal{J}_{1,R}(x,y)$. Indeed, from (\ref{eq8}) we have

\b N_1(x,y,k)&=&
\frac{f^-(x,k)f^+(y,k)}{a(k)}-f^+(x,k)[f^+(y,-k)+\mathcal{R}(k)f^+(y,k)]\nonumber\\
&=&\frac{f^-(x,k)f^+(y,k)}{a(k)}-f^+(y,-k)\frac{[f^-(x,-k)-b(-k)f^+(x,-k)]}{a(-k)}\nonumber
\\&-&\mathcal{R}(k)f^+(x,k)f^+(y,k)\nonumber \\&=&\frac{f^-(x,k)f^+(y,k)}{a(k)}-\frac{f^+(y,-k)f^-(x,-k)}{a(-k)}
\nonumber
\\&-&\mathcal{R}(-k)f^+(x,-k)f^+(y,-k)-\mathcal{R}(k)f^+(x,k)f^+(y,k).\nonumber\e

\n Now it remains to equate the two expressions for
\begin{equation}\label{eq:spec1}
\mathcal{J}_1(x,y) = \frac{1}{2\pi i} \oint_{\gamma_+} R_1(x,y,k)
= \mathcal{J}_{1,\infty} (x,y)+\mathcal{J}_{1,R}(x,y).
\end{equation}
to obtain the completeness relation for the Jost solutions.
\end{proof}

\begin{remark}\label{rem:Cpm}
Another way to prove the completeness relations for the Jost
solutions is to note that $\bar{R}_1(x,y,\bar{k})$ is a kernel of
the resolvent defined for $\im k<0$ and then apply the contour
integration method to the integral
\begin{equation}\label{eq:Jpm}
\mathcal{J}'_1(x,y) = \frac{1}{2\pi i} \oint_{\gamma_+}
\text{d}k\, R_1(x,y,k) - \frac{1}{2\pi i} \oint_{\gamma_-}
\text{d}k\, \bar{R}_1(x,y,\bar{k}).
\end{equation}
\end{remark}

\section{Wronskian relations and generalized Fourier series
expansion}\label{sec:5}

An important tool for the analysis of the mapping between the
potential $m(x)$ of eq. (\ref{eq3}) and the scattering data,
related to this potential, are the so-called {\it Wronskian
relations}.

For what follows, we define the squared eigenfunctions

\begin{equation}\label{eq23}
F^{\pm}(x,k)\equiv (f^{\pm}(x,k))^2, \qquad F^{\pm}_n(x)\equiv
F^{\pm}(x,i\kappa_n).
\end{equation}

\begin{proposition}\label{pr3}
Let $f(k_1,x)$ and $g(k_2,x)$ be two eigenfunctions of the
spectral problem (\ref{eq3}). Then the following identity
(Wronskian relation) holds:

\b \int_{-\infty}^{\infty}q(x)f(k_1,x)g(k_2,x)\text{d}x=
\frac{f_x(k_1,x)g(k_2,x)-f(k_1,x)g_x(k_2,x)}{\lambda(k_1) -
\lambda(k_2)}\Big|_{x=-\infty}^{\infty}\label{eq29} \e
\end{proposition}

\begin{proof} It follows immediately from the fact that $f(k_1,x)$
and $g(k_2,x)$ satisfy (\ref{eq3}).

\begin{cor}
From (\ref{eq29}), (\ref{eq5aa}), (\ref{eq6}), (\ref{eq8}), and
from the formula \b
\lim_{x\rightarrow\infty}\mathrm{P}\frac{e^{ikx}}{k}=\pi i
\delta(k),\nonumber \e

\n where $\mathrm{P}$ means Principal Value, one can obtain
various 'orthogonality' relations for the eigenfunctions, such as
(cf. \cite{CI06})

\b
\int_{-\infty}^{\infty}q(x)f^+(k_1,x)f^-(k_2,x)\text{d}x&=&-2\pi
\omega
a(k_1)\delta(k_1-k_2), \\
\int_{-\infty}^{\infty}q(x)f_{n}^+(x)f_{p}^-(x)\text{d}x&=&i\omega
\dot{a}_n \delta_{np} . \label{eq30} \e
\end{cor} \end{proof}

Let us define a scalar product as usual by

\b (f,g)\equiv
\int_{-\infty}^{\infty}f(x)g(x)\text{d}x,\label{eq31} \e

\n provided the corresponding integral exists.

We need also the skew-symmetric product

\b \biglb f,g \bigrb \equiv
\int_{-\infty}^{\infty}q(x)(f_x(x)g(x)-g_x(x)f(x))\text{d}x,\label{eq31a}
\e

\n related to the Poisson bracket (\ref{PBa}) via \b
\{A,B\}=\Bigglb \frac{\delta A}{\delta m},\frac{\delta B}{\delta
m}\Biggrb .\label{eq31b} \e

Some other useful Wronskian relation are formulated in the next
two propositions.

\begin{proposition}\label{pr4}
Let $f(k,x)$ be an eigenfunctions of the spectral problem
(\ref{eq3}). Then the following identity holds: \b
(m_x,f^2)=-\frac{1}{\lambda}\Big(f_x^2-ff_{xx}\Big)\Big|_{x=-\infty}^{\infty}\label{eq32}
\e
\end{proposition}

\begin{proof}  Since $f(x)$ satisfies (\ref{eq3}), \b
(f_x^2-ff_{xx})\Big|_{x=-\infty}^{\infty} &=&
\int_{-\infty}^{\infty}(f_x^2-ff_{xx})_x \text{d}x=
-\lambda(m_x,f^2).\label{eq33} \e
\end{proof}

The right hand side of (\ref{eq33}) can be expressed also through
the skew-symmetric product (\ref{eq31a}):
\begin{equation}\label{eq:33a}
(m_x,f^2(x,k))\equiv (q_x,f^2(x,k)) = \Bigglb f^2(x,k),
\left(\sqrt{\frac{\omega}{q(x)}}-1\right) \Biggrb.
\end{equation}

\begin{cor}
From Proposition \ref{pr4} and equations (\ref{eq14}),
(\ref{eq14a}) and (\ref{eq:33a}) it follows

\b (m_x,F^{\pm}(k)) \equiv \Bigglb F^\pm(x,k), \left(
\sqrt{\frac{\omega}{q(x)}}-1\right) \Biggrb
=-\frac{4k^2}{\lambda}a(k)b(\mp k)\label{eq34} . \e

\n and the limiting procedure gives on the discrete spectrum

\begin{equation}\label{eq35}
(m_x,F_n^{\pm})= \frac{1}{2} \biglb q(x), F^\pm_n(x)\bigrb
=0,\qquad (m_x,\dot{F}_n^{\pm})= \frac{1}{2} \biglb q(x),
\dot{F}^\pm_n(x)\bigrb = \pm\frac{4i\kappa_n^2}{\lambda_n
C_n^{\pm}},
\end{equation}
\n where $\lambda_n\equiv \lambda(i\kappa_n)$.
\end{cor}

Another type of Wronskian relations relate the variations of the
potential $m(x)$ with the variation of the scattering data
\cite{CI06}:
\begin{equation}\label{eq:var}
\left.    \left( f(x,k)\delta f_x - f_x\delta f(x,k) \right)
\right|_{x=-\infty}^\infty = \int_{-\infty}^\infty \text{d}x\,
\lambda \delta q(x) f^2(x,k),
\end{equation}
where $\delta f(x,k)$ is the variation of the Jost solution
$f(x,k)$ corresponding to the variation $\delta q(x)$ of the
potential. The right hand side of (\ref{eq:var}) can also be
expressed through the skew symmetric  product as follows:
\begin{equation}\label{eq:var-1}
\left.    \left( f(x,k)\delta f_x - f_x\delta f(x,k) \right)
\right|_{x=-\infty}^\infty = -\lambda \biglb f^2(x,k), \delta
\mathcal{Q}_\pm (x) \bigrb ,
\end{equation}
where $\delta \mathcal{Q}_\pm (x) = (1/\sqrt{q(x)}
\int_{\pm\infty}^x \text{d}y \, \delta \sqrt{q(y)}$. For the derivation of (\ref{eq:var-1}) we
performed an integration by parts and used the condition
$\delta \mathcal{Q}_\pm (\mp\infty ) \equiv \pm \delta \alpha =0$.

From (\ref{eq:var}) and (\ref{eq:var-1}) we obtain:
\begin{eqnarray}\label{eq:ss-1}
\biglb F^\mp(x,k) , \delta \mathcal{Q}_\pm \bigrb &=& \mp
\frac{2ik}{\lambda} a^2(k) \delta \mathcal{R}^\pm (k), \\
\label{eq:ss-2} \biglb F^\mp_n(x) , \delta \mathcal{Q}_\pm \bigrb
&=& \frac{2\kappa_n}{\lambda_n C_n^\pm}  \delta \kappa_n, \\
\biglb \dot{F}^\mp_n(x) , \delta \mathcal{Q}_\pm \bigrb &=&
\frac{2i\kappa_n}{\lambda_n}\dot{a}_n^2\delta C_n^\pm -  \left(
\frac{4\kappa_n^2 +1}{4i\omega\lambda_n\kappa_n} -
\frac{\ddot{a}_n}{\dot{a}_n}\right) \frac{\omega \delta
\lambda_n}{\lambda_nC_n^\pm }. \label{eq:ss-3}
\end{eqnarray}

\n Using (\ref{eq:var}) one can also derive the following
relations for the variations of the scattering data, for details
see \cite{CI06}: \b \label{eq37} \frac{\delta
a(k)}{\delta m(x)}&=&-\frac{\lambda}{2ik}f^+(x,k)f^-(x,k)\\
\frac{\delta b(k)}{\delta
m(x)}&=&\frac{\lambda}{2ik}f^+(x,-k)f^-(x,k)\label{eq38}\\\frac{\delta
\ln \lambda_n}{\delta m(x)}&=&\frac{iF_n^-(x)}{\omega b_n
\dot{a}_n}\label{eq39}\e

\n The equations (\ref{eq37}) -- (\ref{eq39}) lead further to

\b \label{eq40} \frac{\delta \mathcal{R}^{\pm}(k)}{\delta
m(x)}&=&\pm\frac{\lambda(k)}{2ika^2(k)}F^{\mp}(x,k)\\
\frac{\delta C_n^{\pm}}{\delta
m(x)}&=&-\frac{\lambda_n}{2i\kappa_n
\dot{a}_n^2}\Big[\Big(\frac{4\kappa_n^2+1}{4i\omega\kappa_n\lambda_n}-
\frac{\ddot{a}_n}{\dot{a}_n}\Big)F_n^{\mp}(x)+\dot{F}_n^{\mp}(x)\Big]\label{eq43}\\
\frac{\delta \kappa_n}{\delta m(x)}&=&-\frac{\lambda_n
C_n^{\pm}}{2\kappa_n}F_n^{\pm}(x)\label{eq41}\\\frac{\delta
\lambda_n}{\delta m(x)}&=&-\frac{1}{\omega}\lambda_n
C_n^{\pm}F_n^{\pm}(x)\label{eq42}\e

These relations allow us to calculate the skew symmetric  products
between the squared solutions.

\begin{proposition}\label{pr4a}
Let $f_{1}(k_1,x)$, $f_{2}(k_2,x)$, $g_{1}(k_1,x)$, $g_{2}(k_2,x)$
be eigenfunctions of the spectral problem (\ref{eq3}). Then

\b \biglb f_1(k_1)g_1(k_1),f_2(k_2)g_2(k_2)\bigrb =
\frac{(f_1\partial_xf_2-f_2\partial_xf_1)
(g_1\partial_xg_2-g_2\partial_xg_1))}
{\lambda(k_1)-\lambda(k_2)}\Big|_{x=-\infty}^{\infty} \nonumber
\\ \label{eq35a} \e
\end{proposition}

\begin{proof}  Straightforward to check using the fact that all
functions are eigenfunctions of the spectral problem (\ref{eq3})
-- see \cite{CI06} for details if necessary. \end{proof}

\begin{cor} From Proposition \ref{pr4a} and (\ref{eq14}), (\ref{eq14a})
it follows

\b \biglb F^{\pm}(k_1),U^{\mp}(k_2)\bigrb =\pm 2\pi \omega
\lambda(k_1)\delta (k_1-k_2), \qquad k_{1,2}\in \mathbb{R}
\label{eq35b} \e

\n where $U^{\pm}(x,k)\equiv
\frac{\lambda(k)}{2ika^2(k)}F^{\pm}(x,k)$. On the discrete
spectrum

\b \biglb \dot{F}^{\pm}_n,F^{\mp}_m\bigrb =\mp 2i \omega \kappa_n
\dot{a}_n^2\delta_{nm}, \qquad \biglb F^{+}_n,F^{-}_m\bigrb
=0,\qquad \biglb \dot{F}^{\pm}_n,F^{\pm}_m \bigrb =0.\label{eq35c}
\e
\end{cor}

Up to now we demonstrated that the mapping between the scattering
data $\mathcal{R}^\pm(k)$ and the potential $m(x)$ (or,
equivalently, $q(x)$) is expressed through the squared solutions
of (\ref{eq3}). The same squared solutions relate also the
variations of the reflection coefficients $\delta
\mathcal{R}^\pm(k)$ with the corresponding variations of the
potential $\delta m(x)$ (or, equivalently, $\delta q(x)$).

In order to ensure that these mappings are one-to-one we have to
prove that the squared solutions form a complete set of functions
in the space of allowed potentials; in other words we will prove
that the functions $F^{\pm}(x, k)$,  $F^{\pm}_n(x)$ form a basis
in the space of allowed potentials. To this end we consider the
function

\b \label{eq25} && R(x,y,k)\equiv
\frac{F^-(x,k)F^+(y,k)}{ka^2(k)}\theta(y-x) \\
&& \qquad +\frac{2f^+(x,k)f^-(x,k)f^+(y,k)f^-(y,k)-
F^-(y,k)F^+(x,k)}{ka^2(k)}\theta(x-y)\nonumber \e

\begin{lemma}\label{lem-2}
\begin{description}
\item [i)] $R(x,y,k)$ is an analytic function of $k$ for $k\in
\bbbc_+$; \item [ii)] $R(x,y,k)$ has  second order poles at
$k=i\kappa_n$; \item [iii)] $R(x,y,k)$ is a kernel of bounded
integral operator for $\im k>0$. For $\im k=0$ $R_1(x,y,k)$ is a
kernel of a unbounded integral operator.
\end{description}
\end{lemma}

\begin{proof} i) and ii) are obvious.

iii) For $\im k>0$ the statement follows from the definitions
(\ref{eq6}) and (\ref{eq6'}) of the Jost solutions, which ensure
that $R(x,y,k)$ falls off exponentially for all $x,y\to\pm
\infty$. The same arguments for $\im k=0$ can only ensure that
$R(x,y,k)$ is bounded function for $x,y \to \pm \infty$.
\end{proof}

\begin{proposition}\label{pr2} The following completeness relation  holds:
\b \frac{\omega}{\sqrt{q(x)q(y)}} \theta(x-y)=-\frac{1}{2\pi i}
\int_{-\infty}^{\infty}\frac{F^-(x,k)F^+(y,k)}{ka^2(k)}\text{d}k
\phantom{**********}\nonumber\\ + \sum_{n=1}^{N}\frac{1}{i\kappa_n
\dot{a}_n^2}\Big[\dot{F}_n^-(x)F_n^+(y)+F_n^-(x)\dot{F}_n^+(y)-
\Big(\frac{1}{i\kappa_n}+\frac{\ddot{a}_n}{\dot{a}_n}\Big)
F_n^-(x)F_n^+(y)\Big].\label{eq24}\e
\end{proposition}

\begin{proof} The proof is similar to the one of Proposition
\ref{pr1}.

Consider the contour integral
\begin{equation}\label{eq:J}
\mathcal{J}(x,y) = \frac{1}{2\pi i} \oint_{\gamma_+} R(x,y,k)
\text{d}k.
\end{equation}
where the contour $\gamma_+$ is shown on Fig. \ref{fig:1}. From
the Cauchy residue theorem it follows that:
\begin{eqnarray}\label{eq:J-res2}
&& \mathcal{J}(x,y) = \sum_{n=1}^N \Res_{k=i\kappa_n} R(x,y,k)
 \\ && \quad =  \sum_{n=1}^{N}\frac{1}{i\kappa_n
\dot{a}_n^2} \Big[\dot{F}_n^-(x)F_n^+(y)+F_n^-(x)\dot{F}_n^+(y)-
\Big(\frac{1}{i\kappa_n}+\frac{\ddot{a}_n}{\dot{a}_n}\Big)
F_n^-(x)F_n^+(y)\Big].\nonumber
\end{eqnarray}

In this evaluation we used (\ref{eq:a-n}), (\ref{eq200}) and the
fact that $\theta(x-y) + \theta(y-x)=1$.

Next we evaluate the integral $\mathcal{J}(x,y)$ by integrating
along the contour. For the integration along the infinite
semicircle we need the asymptotic of $R(x,y,k)$ for
$|k|\rightarrow \infty$. Due to  (\ref{eq21aa}) and (\ref{eq22})
we get \b\label{eq26}
R(x,y,k)&=&\frac{2}{k}\frac{\omega}{\sqrt{q(x)q(y)}}\theta(x-y)
(1+o(1/k)) \\
&+& \frac{1}{k} \left( R_0(x,y,k)\theta(y-x) -
R_0(x,y,-k)\theta(x-y) \right)(1+o(1/k)), \nonumber  \\
R_0(x,y,k) &=& \frac{\omega}{\sqrt{q(x)q(y)}} e^{2ik\int_x^y
\text{d}y \sqrt{q(y)/\omega}}. \nonumber  \e

Only the leading terms in (\ref{eq260}), which are are entire
functions of $k$ contribute to the integral. This allows us to
deform the infinite semicircle until it coincides with the real
$k$-axis. Then the integration over $k$ is easily performed with
the result:
\begin{equation}\label{eq:J80}
\mathcal{J}_{\infty} (x,y) =  \frac{\omega}{\sqrt{q(x)q(y)}}
\theta(x-y).
\end{equation}

To evaluate the integral over the real axis $\mathcal{J}_{R}(x,y)$
we will use the the fact that $R(x,y,k)$ can be written in the
form \b R(x,y,k)&=& \frac{F^-(x,k)F^+(y,k)}{k a(k)} -
\frac{N_1^2(x,y,k)}{k} \theta(x-y) \label{eq270}\e where
$N_1(x,y,k)$ is defined in (\ref{eq280}). The second term in
(\ref{eq270-1}) is an odd function of $k$ and does not contribute
to $\mathcal{J}_{R}(x,y)$.

Now it remains to equate the two expressions for
\begin{equation}\label{eq:spec}
\mathcal{J}(x,y) = \frac{1}{2\pi i} \oint_{\gamma_+}
R(x,y,k)\text{d} k = \mathcal{J}_{\infty}
(x,y)+\mathcal{J}_{R}(x,y).
\end{equation}
to obtain the completeness relation for the squared solutions.
\end{proof}

\begin{cor}\label{cor:2}
The completeness relation (\ref{eq24}) can be rewritten in the
following equivalent form: \b && \theta(x-y)=-\frac{1}{2\pi i}
\int_{-\infty}^{\infty}\frac{\tilde{F}^-(x,k)\tilde{F}^+(y,k)}
{ka^2(k)}\text{d}k \label{eq24'} \\ && \quad +
\sum_{n=1}^{N}\frac{1}{i\kappa_n
\dot{a}_n^2}\Big[\dot{\tilde{F}}_n^-(x)\tilde{F}_n^+(y)+
\tilde{F}_n^-(x)\dot{\tilde{F}}_n^+(y)-
\Big(\frac{1}{i\kappa_n}+\frac{\ddot{a}_n}{\dot{a}_n}\Big)
\tilde{F}_n^-(x)\tilde{F}_n^+(y)\Big]. \nonumber \e where
\begin{equation}\label{eq:t-F}
\tilde{F}^\pm (x,k) = \sqrt{\frac{q(x)}{\omega}} F^\pm(x,k).
\end{equation}
\end{cor}

The completeness relation allows one to expand any function $X(x)$
over the squared solutions $F^+(x,k)$ or $F^-(x,k)$. To this end
we multiply both sides of (\ref{eq24}) with $\frac{1}{2} q_y X(y)
+ q(y)X_y$ (resp. with $\frac{1}{2} q_x X(x) + q(x)X_x$) and
integrate over $\text{d}y$ (resp. $\text{d}x$). A simple
calculation gives:

\begin{equation}\label{eq:X'}
 \pm \omega  X(x) = -\frac{1}{2\pi i} \int_{-\infty}^\infty
\frac{\text{d}k}{k}
\frac{F^\mp (x,k)  \xi_X(k)}{a^2(k)}  
+ \sum_{n=1}^N \frac{1}{i\kappa_n \dot{a}_n^2} \left( \dot{F}^\mp
\xi^\pm_{X,n} +F^\mp \dot{\xi}^\pm_{X,n} \right),
\end{equation}
provided $X(x)$ vanishes for $x\to\pm\infty$.  The expansion
coefficients $\xi_X^\pm(k)$ are given by:
\begin{equation}\label{eq:X-1}
\xi_X^\pm (k) = \int_{-\infty}^\infty \text{d}k F^\pm (y,k)\left(
\frac{1}{2} q_y X(y) +q(y)X(y)\right)  = \frac{1}{2}\biglb F^\pm
(x,k),X(x)\bigrb,
\end{equation}
where $\biglb \cdot\, , \cdot \, \bigrb $ is the skew-symmetric
product (\ref{eq31a}). Similarly for  $\xi_{X,n}^\pm$ and
$\dot{\xi}_{X,n}^\pm$ we find:
\begin{equation}\label{eq:X-1d}
\xi_{X,n}^\pm = \frac{1}{2}\biglb F^\pm _n(x),X(x) \bigrb, \qquad
\dot{\xi}^\pm_{X,n} = \frac{1}{2}\biglb \dot{F}_n^\pm
(x),X(x)\bigrb - \left( \frac{1}{i\kappa_n} +
\frac{\ddot{a}_n}{\dot{a}_n} \right) \xi_{X,n}^\pm,
\end{equation}

Analogously, if we multiply the completeness relation
(\ref{eq24'}) by $\tilde{X}_y$ (resp. by  $\tilde{X}_x$) and
integrate over $\text{d}y$ (resp. $\text{d}x$) we get:
\begin{equation}\label{eq:X}
\pm \omega  \tilde{X}(x) = -\frac{1}{2\pi i} \int_{-\infty}^\infty
\frac{\text{d}k}{k} \frac{\tilde{F}^\mp (x,k)
\tilde{\xi}_{\tilde{X}}(k)}{a^2(k)} + \sum_{n=1}^N
\frac{1}{i\kappa_n \dot{a}_n^2} \left( \dot{\tilde{F}}^\mp
\tilde{\xi}^\pm_{\tilde{X},n} + \tilde{F}^\mp
\dot{\tilde{\xi}}^\pm_{\tilde{X},n}  \right),
\end{equation}
where the expansion coefficients $\tilde{\xi}_{\tilde{X}}^\pm(k)$,
$\tilde{\xi}_{\tilde{X},n}^\pm$ and
$\dot{\tilde{\xi}}_{\tilde{X},n}^\pm$ are expressed by another
skew-symmetric product:
\begin{eqnarray}\label{eq:X-1'}
&& \tilde{\xi}_{\tilde{X}}^\pm (k) = \int_{-\infty}^\infty
\text{d}k \tilde{F}^\pm (y,k) \tilde{X}_y(y) = \frac{1}{2}\biglb
\tilde{F}^\pm (x,k), \tilde{X}(x)\bigrbt, \\
&& \tilde{\xi}_{\tilde{X},n}^\pm = \frac{1}{2}\biglb
\tilde{F}_n^\pm (x), \tilde{X}(x)\bigrbt , \qquad
\dot{\tilde{\xi}}^\pm_{\tilde{X},n} = \frac{1}{2}\biglb
\tilde{F}_n^\pm (x), \tilde{X}(x)\bigrbt  - \left(
\frac{1}{i\kappa_n} + \frac{\ddot{a}_n}{\dot{a}_n} \right)
\tilde{\xi}_{\tilde{X},n}^\pm, \nonumber
\end{eqnarray}

Obviously the two skew-symmetric products are related by a
gauge-like transformation. Indeed, if $F^\pm(x,k)$ and
$\tilde{F}^\pm(x,k)$ are related by (\ref{eq:t-F}) and if
$\tilde{X}(x)= \sqrt{q(x)/\omega} X(x) $ then:
\begin{equation}\label{skew-pr}
\biglb F^\pm(x,k), X(x)\bigrb =\omega \biglb \tilde{F}^\pm(x,k),
    \tilde{X}(x)\bigrbt.
\end{equation}

The following lemma demonstrates that the mapping between $X(x)$
and its set of expansion coefficients is one-to-one.

\begin{lemma}\label{lem:2}
\begin{description}
\item[i)] A necessary and sufficient condition for $X(x)$ to
vanish is that all its expansion coefficients vanish:
\begin{equation}\label{eq:ex-c0}
\xi_X^\pm(k)=0, \qquad \xi_{X,n}^\pm=0,\qquad \dot{\xi}_{X,n}^\pm
=0.
\end{equation}
\item[ii)] A necessary and sufficient condition for $\tilde{X}(x)$
to vanish is:
\begin{equation}\label{eq:ex-c1}
\tilde{\xi}_{\tilde{X}}^\pm(k)=0, \qquad \tilde{\xi}_{\tilde{X},n}
^\pm=0,\qquad \dot{\tilde{\xi}}_{\tilde{X},n}^\pm =0.
\end{equation}
\end{description}
\end{lemma}

\begin{proof}
i) Let us assume that $X(x)=0$ for all $x$. Substituting it into
the right hand side of (\ref{eq:X-1}),  (\ref{eq:X-1d}) we
immediately get (\ref{eq:ex-c0}). Let us now assume that
 (\ref{eq:ex-c0}) holds. Inserting it into the right hand side of
the expansion (\ref{eq:X'}) we find that $X(x) =0$ for all $x$.
ii) is proved analogously using (\ref{eq:X-1'}) and the expansion
(\ref{eq:X}).
\end{proof}

\begin{remark}
We introduced two sets of squared solutions $F^\pm(x,k)$ and
$\tilde{F}^\pm(x,k)$ related by the gauge-like transformation
(\ref{eq24'}). Since $q(x)$ is always positive the two sets are
obviously equivalent.
\end{remark}

\section{Symplectic basis and canonical variables}\label{sec:6}

Here we introduce a third set of squared solutions, known as {\it
symplectic basis} whose special properties will become clear
below. The symplectic basis was introduced for the first time in
analyzing the nonlinear Schr\"odinger (NLS) equation in \cite{GeHr1}.

The canonical (action-angle) variables for the CH equation are
known \cite{CI06}: \b \rho(k)&\equiv &\frac{2k}{\pi \omega
\lambda(k)^2}\ln |a(k)|=-\frac{k}{\pi \omega \lambda(k)^2}
\ln(1-\mathcal{R}^{\pm}(k)\mathcal{R}^{\pm}(-k)), \label{eq52} \\
\phi (k)&\equiv &\arg b(k)=\frac{1}{2i}\ln
\frac{\mathcal{R}^+(k)}{\mathcal{R}^-(k)}, \label{eq53} \\
\rho_n&=&\lambda_n^{-1}, \qquad \phi_n=\ln |b_n|, \label{eq54} \e

\n satisfying \b  \{\rho(k_1),\phi(k_2)\}&=&\delta(k_1-k_2)\label{eq55}\\
\{\rho(k_1),\rho(k_2)\}&=&\{\phi(k_1),\phi(k_2)\}=0, \qquad
k_{1,2}>0 \label{eq56} \\
\{\rho_m,\phi_n\}&=&\delta_{mn}, \label{eq57} \\
\{\rho_m,\rho_n\}&=&\{\phi_m,\phi_n\}=0.\label{eq58}
 \e

The following proposition gives the symplectic basis in the space
of Schwartz-class functions:
\begin{proposition}\label{pr9}
The quantities
\b \mathcal{P}(x,k)&=&\frac{1}{2\pi i \omega \lambda(k)} \Big(\mathcal{R}^-(k)F^-(x,k)-\mathcal{R}^-(-k)F^-(x,-k)\Big)\nonumber \\
&=&-\frac{1}{2\pi i \omega \lambda(k)} \Big(\mathcal{R}^+(k)F^+(x,k)-\mathcal{R}^+(-k)F^+(x,-k)\Big),\label{eq59}\\
\mathcal{Q}(x,k)&=&-\frac{\lambda(k)}{4kb(k)b(-k)} \Big(\mathcal{R}^-(k)F^-(x,k)+\mathcal{R}^+(k)F^+(x,k)\Big),\label{eq60} \\
P_n(x)&=&\frac{1}{\omega \lambda_n}C_n^+F_n^+(x)=\frac{1}{\omega
\lambda_n}C_n^-F_n^-(x),\label{eq61}\\Q_n(x)&=&
\frac{i\lambda_n}{4\kappa_n}\Big(C_n^+\dot{F}_n^+(x)-C_n^-\dot{F}_n^-(x)\Big).\label{eq61a}\e
satisfy the following canonical relations: \b
\biglb \mathcal{P}(k_1),\mathcal{Q}(k_2)\bigrb &=&\delta(k_1-k_2),\label{eq62}\\
\phantom{*}\biglb \mathcal{P}(k_1),\mathcal{P}(k_2) \bigrb
&=&\biglb \mathcal{Q}(k_1),\mathcal{Q}(k_2)\bigrb =0, \qquad
k_{1,2}>0 \label{eq63} \\ \phantom{*} \biglb P_m,Q_n\bigrb
&=&\delta_{mn}, \label{eq64}
\\ \phantom{*} \biglb P_m,P_n\bigrb &=& \biglb Q_m,Q_n \bigrb =0.\label{eq65}
 \e
\end{proposition}

\begin{proof}  Due to (\ref{eq31b}), \b
\mathcal{P}(x,k)&=&\frac{\delta\rho(k)}{\delta m(x)},\qquad
\mathcal{Q}(x,k)=\frac{\delta\phi(k)}{\delta m(x)}, \qquad
\label{eq66a} \\ P_n(x)&=&\frac{\delta\rho_n}{\delta m(x)},\qquad
Q_n(x)=\frac{\delta\phi_n}{\delta m(x)} \label{eq66b} \e

\n and with (\ref{eq40}) -- (\ref{eq42}) we obtain (\ref{eq59}) --
(\ref{eq61}). From (\ref{eq53}), (\ref{eq54}) one can see that
$\phi_n$ can be obtained by a limiting procedure $k\rightarrow
i\kappa_n$ from $\phi(k)$ (up to an overall multiplier $i$).
Therefore, $Q_n(x)$ can be obtained in the same way from $Q(x,k)$,
giving (\ref{eq61a}). Note that (\ref{eq64}) is satisfied due to
(\ref{eq35c}), (\ref{eq61}) and (\ref{eq61a}). \end{proof}

The usefulness of the symplectic basis (\ref{eq59}) --
(\ref{eq61}) is more evident from the following two propositions.
\begin{proposition}\label{pr10}
\b   \frac{1}{2\sqrt{q(x)q(y)}}\varepsilon (x-y) &=&
\int_{0}^{\infty}\Big(\mathcal{P}(x,k)\mathcal{Q}(y,k)-
\mathcal{Q}(x,k),\mathcal{P}(y,k)\Big)dk\nonumber \\
&+& \sum_{n=1}^{N}\Big(P_n(x)Q_n(y)-Q_n(x)P_n(y)\Big).
\label{eq63a}\e where $\varepsilon (x-y) = \Big(
\theta(x-y)-\theta(y-x)\Big)/2$.
\end{proposition}
\begin{proof}  From (\ref{eq59}) we have the identity

\b \mathcal{R}^-(k)F^-(x,k)+\mathcal{R}^+(k)F^+(x,k)&=&\nonumber
\\ \mathcal{R}^-(-k)F^-(x,-k)&+&\mathcal{R}^+(-k)F^+(x,-k).
\label{eq64a} \e

\n Applying several times (\ref{eq64a}) one can check further that

\b 4\omega\Big(P(x,k)Q(y,k)-Q(x,k)P(y,k)\Big)=\phantom{*************} \nonumber \\
-\frac{1}{2\pi
i}\Big(\frac{F^-(x,k)F^+(y,k)-F^+(x,k)F^-(y,k)}{ka^2(k)}+(k\rightarrow
-k)\Big) \label{eq65a}  \e

\n With (\ref{eq61}) and (\ref{eq61a}) it is straightforward to
obtain

\b
4\omega\sum_{n=1}^{N}\Big(P_n(x)Q_n(y)-Q_n(x)P_n(y)\Big)=\phantom{*************}
\nonumber
\\\sum_{n=1}^{N}\Big(\frac{\dot{F}_n^-(x)F_n^+(y)-\dot{F}_n^+(x)F_n^-(y)}{i\kappa_n
\dot{a}_n^2}+(x\leftrightarrow y)\Big). \label{eq66}  \e

\n Now, (\ref{eq63a}) follows from (\ref{eq65a}), (\ref{eq66}) and
the representation (\ref{eq24}). \end{proof}

The completeness relation (\ref{eq66a}) allows one to expand any
smooth  function $X(x)$ vanishing for $x\to\pm\infty$ over the
symplectic basis. To this end we multiply both sides of
(\ref{eq66a}) by $\frac{1}{2} q_yX(y) + q(y)X_y$ and integrate
over $\text{d}y$. The result is:
\begin{equation}\label{eq:X-PQ}
    X(x)= \int_0^\infty \text{d}k\, \left( \mathcal{P}(x,k)
    \phi_X(k) - \mathcal{Q}(x,k)\rho_X(k)\right) + \sum_{n=1}^N
    \left( P_n(x)\phi_{n,X} - Q_n(x)\rho_{n,X}\right),
\end{equation}
where
\begin{eqnarray}\label{eq:X-phi}
\phi_X(k)  &=& \biglb \mathcal{Q}(y,k),X(y)\bigrb , \qquad
\rho_X(k) =\biglb \mathcal{P}(y,k),X(y)\bigrb , \nonumber \\
\phi_{n,X} &=& \biglb Q_n(y),X(y)\bigrb , \qquad \rho_{n,X}
=\biglb P_n(y),X(y)\bigrb .
\end{eqnarray}

\begin{cor}\label{cor:PQ-t}
The completeness relation (\ref{eq63a}) can be cast in the
following equivalent form: \b  && \frac{1}{2}\varepsilon (x-y) =
\int_{0}^{\infty}\Big(\tilde{\mathcal{P}}(x,k)
\tilde{\mathcal{Q}}(y,k)- \tilde{\mathcal{Q}}(x,k),
\tilde{\mathcal{P}}(y,k)\Big)dk\nonumber \\
&+& \sum_{n=1}^{N}\Big(\tilde{P}_n(x)
\tilde{Q}_n(y)-\tilde{Q}_n(x) \tilde{P}_n(y)\Big), \label{eq63a'}
\e where the elements of the new symplectic basis
$\tilde{\mathcal{P}}(x,k)$, $\tilde{\mathcal{Q}}(x,k)$ are related
to the old ones by:
\begin{equation}\label{eq:P-P-t}
\tilde{\mathcal{P}}(x,k) = \sqrt{q(x)}\mathcal{P}(x,k), \qquad
\tilde{\mathcal{Q}}(x,k) = \sqrt{q(x)}\mathcal{Q}(x,k),
\end{equation}
for all $k\in \bbbr \cup \{i\kappa_n\} $.
\end{cor}

With (\ref{eq63a'}) the analogue of the expansion (\ref{eq:X-PQ})
for any smooth function $\tilde{X}(x)$ vanishing for
$x\to\pm\infty$ is:
\begin{eqnarray}\label{eq:X-s}
&& \tilde{X}(x) =  \int_{-\infty}^\infty \text{d}k
\left(\tilde{\mathcal{P}}(x,k) \tilde{\phi}_X(k) -
\tilde{\mathcal{Q}}(x,k) \tilde{\rho}_X(k) \right)
\nonumber \\
&+& \sum_{n=1}^N \left( \tilde{P}_n \tilde{\phi}_{X,n} -
\tilde{Q}_n \tilde{\rho}_{X,n} \right),
\end{eqnarray}
where the expansion coefficients $\tilde{\rho}_X(k)$,
$\tilde{\rho}_{X,n}$ and $\tilde{\phi}_X(k)$, $\tilde{\phi}_{X,n}$
are given by:
\begin{eqnarray}\label{eq:X-sb}
\tilde{\rho}_X(k) &=& \biglb \tilde{\mathcal{Q}}(k) , \tilde{X}
\bigrbt , \qquad \tilde{\phi}_X(k) = \biglb \tilde{\mathcal{P}}(k)
, \tilde{X} \bigrbt, \\
\tilde{\rho}_{X,n}&=& \biglb \tilde{Q}_n , \tilde{X} \bigrbt ,
\qquad \tilde{\phi}_X(k) = \biglb \tilde{P}_n, \tilde{X}\bigrbt ,
\nonumber
\end{eqnarray}
where $\biglb \; \cdot , \; \cdot \bigrbt$ is the new
skew-symmetric product (\ref{skew-pr}).

The following lemma demonstrates that the mapping between
$\tilde{X}(x)$ and its set of expansion coefficients is
one-to-one.

\begin{lemma}\label{lem:2'}
A necessary and sufficient condition for $\tilde{X}(x)$ to vanish
identically is that all its expansion coefficients vanish:
\begin{equation}\label{eq:ex-c0-}
\tilde{\rho}_{\tilde{X}}(k) =0, \qquad \tilde{\phi}_{\tilde{X}}
(k), \qquad \tilde{\rho}_{\tilde{X},n} =0, \qquad
\tilde{\phi}_{\tilde{X},n}=0.
\end{equation}
\end{lemma}
The proof is analogous to the one of Lemma \ref{lem:2}.


\section{Recursion operator for the Camassa-Holm
hierarchy}\label{sec:7}

As in Section \ref{sec:5} we can view the completeness relations
(\ref{eq24}) and (\ref{eq24'}) as spectral decompositions for the
recursion operators $L_\pm$ and $\tilde{L}_\pm$ defined below.

\begin{proposition}\label{pro:L-L} Let us define the recursion operators
$L_\pm $ and $\tilde{L}_\pm$ and their inverse $\hat{L}_\pm $ and
$\hat{\tilde{L}}_\pm$ as follows: \b
L_{\pm}=\mathcal{D}_x^{-1}\Big[4q(x)-2\int_{\pm
\infty}^{x}\text{d}y \, m'(y)\Big].\label{eq44} \qquad
\tilde{L}_\pm = \frac{4\omega}{\sqrt{q}}  \mathcal{D}_x^{-1}
\int_{\pm\infty}^x \text{d} y\, \sqrt{q(y)} \partial_x \cdot, \\
\hat{L}_\pm =\frac{1}{4\sqrt{q}} \int_{\pm\infty}^x \text{d} y \,
\frac{1}{\sqrt{q}} \mathcal{D}_y \partial_y\, \cdot , \qquad
\hat{\tilde{L}}_\pm = \int_{\pm\infty}^x \text{d} y\,
\frac{1}{\sqrt{q}} \partial_y \mathcal{D}_y
\frac{\sqrt{q}}{4\omega} \cdot. \label{eq46i}
 \e
where $\mathcal{D}_x=\partial_x^2-1$. Then the following relations
hold:

\b L_{\pm}F^{\pm}(x,k)&=&\frac{1}{\lambda}F^{\pm}(x,k), \qquad
L_{\pm}F^{\pm}_n(x)=\frac{1}{\lambda_n}F^{\pm}_n(x), \label{eq45}
\\ L_{\pm}\dot{F}^{\pm}_n(x)&=&\frac{1}{\lambda_n}\dot{F}^{\pm}_n(x) +
\frac{2ik_n}{\omega\lambda_n^2} F_n^\pm(x). \nonumber\\
\tilde{L}_{\pm}\tilde{F}^{\pm}(x,k)&=&
\frac{1}{\lambda}\tilde{F}^{\pm}(x,k), \qquad
\tilde{L}_{\pm}\tilde{F}^{\pm}_n(x)=
\frac{1}{\lambda_n}\tilde{F}^{\pm}_n(x), \label{eq45'}
\\ \tilde{L}_{\pm}\dot{\tilde{F}}^{\pm}_n(x)&=&
\frac{1}{\lambda_n}\dot{\tilde{F}}^{\pm}_n(x) +
\frac{2ik_n}{\omega\lambda_n^2} \tilde{F}_n^\pm(x). \nonumber \\
\hat{L}_{\pm}F^{\pm}(x,k)&=&\lambda F^{\pm}(x,k), \qquad
\hat{L}_{\pm}F^{\pm}_n(x)=\lambda_nF^{\pm}_n(x), \label{eq45i}
\\ \hat{L}_{\pm}\dot{F}^{\pm}_n(x)&=&\lambda_n\dot{F}^{\pm}_n(x) -
\frac{2ik_n}{\omega} F_n(x)^\pm. \nonumber\\
\hat{\tilde{L}}_{\pm}\tilde{F}^{\pm}(x,k)&=&
\lambda\tilde{F}^{\pm}(x,k), \qquad
\hat{\tilde{L}}_{\pm}\tilde{F}^{\pm}_n(x)=
\lambda_n\tilde{F}^{\pm}_n(x), \label{eq45i'}
\\ \hat{\tilde{L}}_{\pm}\dot{\tilde{F}}^{\pm}_n(x)&=&
\lambda_n\dot{\tilde{F}}^{\pm}_n(x) - \frac{2ik_n}{\omega}
\tilde{F}_n^\pm(x). \nonumber \e
\end{proposition}

\begin{proof}
It is not difficult to prove that $L_\pm \hat{L}_\pm$ and
$\tilde{L}_\pm \hat{\tilde{L}}_\pm$ act as an identity operator on
any function $X(x)$. In order to prove (\ref{eq45}) one can make
use of the fact that it can be reformulated as follows: \b
\Big[\partial_{x}^{2}-4\lambda q(x)+ 2 \lambda \int_{\pm
\infty}^{x}\text{d}y \, m'(y) \bullet \Big]
(f^{\pm})^2=(f^{\pm})^2,  \label{eq46}
 \e
This follows from the fact that $f^\pm(x,k)$ are eigenfunctions of
the spectral problem (\ref{eq3}) with the asymptotics (\ref{eq6})
and (\ref{eq6'}). The rest of the relations are proved
analogously.
\end{proof}

\begin{cor}
The recursion operators $L_\pm$, $\tilde{L}_\pm$ and their inverse
$\hat{L}_\pm$, $\hat{\tilde{L}}_\pm$ satisfy the following
relations:
\begin{eqnarray}\label{eq:L-ssp}
\biglb L_+X , Y\bigrb = \biglb X , L_-Y\bigrb, \qquad     \biglb
\hat{L}_+X , Y\bigrb = \biglb X ,    \hat{L}_-Y\bigrb, \\
\biglb \tilde{L}_+X , Y\bigrbt = \biglb X , \tilde{L}_-Y\bigrbt,
\qquad \biglb \hat{\tilde{L}}_+X , Y\bigrbt = \biglb X ,
\hat{\tilde{L}}_-Y\bigrbt,
\end{eqnarray}
for any pair of functions $X(x)$ and $Y(x)$.
\end{cor}

\begin{proof} It follows easily from the definitions of the
skew symmetric  product and of the recursion operators  using
integration by parts.

\end{proof}

Let us now define the operator $L\equiv \frac{1}{2}(L_+ + L_-)$
cf. (\ref{eq44}), and $\tilde{L}\equiv \frac{1}{2}(\tilde{L}_+ +
\tilde{L}_-)$. A simple direct computation shows that a kernel of
$L$ for $\omega\ne0$ is empty, therefore it is possible to define
the inverse operator $L^{-1}$. It is clear from (\ref{eq59}),
(\ref{eq61}), (\ref{eq45}) that

\b \label{eq:LP2} L \mathcal{P}(x,k)=\lambda^{-1}\mathcal{P}(x,k),
\qquad L P_n(x)=\lambda_n ^{-1 } P_n(x)  .\e

One can compute the action of $L$ to the remaining part of the
symplectic basis, making use of the following proposition:

\begin{proposition}\label{pr11aa}
The following relations hold:

\b \label{eq:mixedsigns} L_{\mp}
F^{\pm}(x,k)=\frac{1}{\lambda}F^{\pm}(x,k) \pm
\frac{8k^2}{\lambda}a(k)b(\mp k).\e
\end{proposition}

\begin{proof}  One can write (\ref{eq:mixedsigns}) as

\b \Big[\partial_{x}^{2}-1-4\lambda q(x)+ 2 \lambda \int_{\mp
\infty}^{x}\text{d}y \phantom{*} m'(y) \bullet \Big] (f^{\pm})^2
\mp \frac{8k^2}{\lambda}a(k)b(\mp k)=0,\nonumber \e

\n which is fulfilled due to the fact that $f^{\pm}$ are
eigenfunctions of the spectral problem (\ref{eq3}) with
asymptotics (\ref{eq6}), (\ref{eq5aa}), (\ref{eq14}),
(\ref{eq14a}). \end{proof}

\begin{cor}
From the above Proposition and (\ref{eq60}) it follows \b
\label{eq:LQ} L_{\pm}
\mathcal{Q}(x,k)=\frac{1}{\lambda}\mathcal{Q}(x,k) \pm 2k .\e
\end{cor}

Now it is clear that the eigenfunctions for the operators $L$ and
$\tilde{L}$ are the elements of the symplectic bases
$\mathcal{P}(x,k)$, $\mathcal{Q}(x,k)$, $P_n(x)$, $Q_n(x)$  and
$\tilde{\mathcal{P}}(x,k)$, $\tilde{\mathcal{Q}}(x,k)$,
$\tilde{P}_n(x)$, $\tilde{Q}_n(x)$, e.g.
\begin{eqnarray}\label{eq:L-P}
L\mathcal{P}(x,k) =\frac{1}{\lambda } \mathcal{P}(x,k), \qquad
L\mathcal{Q}(x,k) =\frac{1}{\lambda } \mathcal{Q}(x,k),\\
\label{eq:L-Pn} LP_n(x) =\frac{1}{\lambda_n } P_n(x), \qquad
LQ_n(x) =\frac{1}{\lambda_n } Q_n(x).
\end{eqnarray}

\begin{proposition}\label{pro:L-L-t}
The inverse of the recursion operators $L$ and $\tilde{L}$ are
given by:
\begin{eqnarray}\label{eq:L-tL}
L^{-1}\equiv\hat{L}= \frac{1}{2} (\hat{L}_++\hat{L}_-), \qquad
\tilde{L}^{-1}\equiv \hat{\tilde{L}}= \frac{1}{2}
(\hat{\tilde{L}}_++\hat{\tilde{L}}_-). \nonumber
\end{eqnarray}

\end{proposition}

\begin{proof} Checked by direct calculation.
\end{proof}

\begin{cor}\label{L-L-ta}
The recursion operators $L$, $\tilde{L}$ and their inverse
$\hat{L}$, $\hat{\tilde{L}}$ are `self-adjoint' with respect to
the skew symmetric  product:
\begin{eqnarray}\label{eq:L-ssp-}
\biglb L X , Y\bigrb = \biglb X , LY\bigrb, \qquad     \biglb
\hat{L}X , Y\bigrb = \biglb X ,    \hat{L}Y\bigrb, \\
\biglb \tilde{L}X , Y\bigrbt = \biglb X , \tilde{L}Y\bigrbt,
\qquad \biglb \hat{\tilde{L}}X , Y\bigrbt = \biglb X ,
\hat{\tilde{L}}Y\bigrbt,
\end{eqnarray}
for any functions $X(x)$ and $Y(x)$.
\end{cor}

\section{Expansions over the squared solutions and the CH hierarchy} \label{sec:8}

It has been demonstrated that the squared solutions satisfy the
completeness relation and therefore can be considered as
generalized exponents. In this section we will derive the
expansions of three important functions and demonstrate how they
can be used for establishing the fundamental properties of the
Camassa-Holm hierarchy.

The first of these functions is the 'potential' $m(x)$, or, rather
one of the functions $\sqrt{\omega/q(x)}-1$ or
$\sqrt{q(x)/\omega}-1$, which are completely determined by $m(x)$
and vice-versa. Its generalized Fourier coefficients are
determined by the scattering data.

\begin{proposition}\label{pr5}
\begin{equation}\label{eq36}
\omega\Big(\sqrt{\frac{\omega}{q(x)}}-1\Big)= \pm\frac{1}{2\pi
i}\int_{-\infty}^{\infty}\frac{2k \mathcal{R}^{\pm}(k)}
{\lambda(k)} F^{\pm}(x,k)\text{d}k +
\sum_{n=1}^{N}\frac{2\kappa_n}{\lambda_n}C_n^{\pm}F_n^{\pm}(x),
\end{equation}
\begin{equation}\label{eq36b}
\omega\Big(1- \sqrt{\frac{q(x)}{\omega}}\Big) = \pm\frac{1}{2\pi
i}\int_{-\infty}^{\infty}\frac{2k
\mathcal{R}^{\pm}(k)}{\lambda(k)} \tilde{F}^{\pm}(x,k)\text{d}k +
\sum_{n=1}^{N}\frac{2\kappa_n}{\lambda_n}C_n^{\pm}
\tilde{F}_n^{\pm}(x),
\end{equation}
\begin{equation}\label{eq72aa}
\sqrt{\frac{\omega}{q(x)}}-1 =-2\int_{0}^{\infty}k\mathcal{P}(x,k)
\text{d}k +2\sum_{n=1}^{N}\kappa_nP_n (x),
\end{equation}
\begin{equation}\label{eq72aa'}
1- \sqrt{\frac{q(x)}{\omega}} = -2\int_{0}^{\infty}k
\tilde{\mathcal{P}}(x,k)\text{d}k +2\sum_{n=1}^{N}\kappa_n
\tilde{P}_n (x).
\end{equation}
\end{proposition}

\begin{proof} The first expansion (\ref{eq36}) is obtained by
multiplying both sides of (\ref{eq24}) by $m_y$, integrating with
respect to $y$ and using (\ref{eq34}), (\ref{eq35}). The second
one (\ref{eq36b})  follows from the first one and from
(\ref{eq:t-F}). The expansion coefficients of the third expansion
(\ref{eq72aa}) can be calculated using (\ref{eq:X-phi}), the
definition of the symplectic basis (\ref{eq59})--(\ref{eq61a}) and
(\ref{eq34}), (\ref{eq35}).  The fourth expansion (\ref{eq72aa'})
is an immediate consequence of the third (\ref{eq72aa}) and
(\ref{eq:P-P-t}).
\end{proof}

The generalized Fourier expansion for the variation of the
'potential' reads as follows:

\begin{proposition}\label{pr6}
\b \frac{\omega}{\sqrt{q(x)}}\int_{\pm
\infty}^{x}\delta\sqrt{q(y)}\text{d}y & = & \frac{1}{2\pi
i}\int_{-\infty}^{\infty}\frac{i}{\lambda(k)}\delta
\mathcal{R}^{\pm}(k)F^{\pm}(x,k)\text{d}k \label{eq43a}  \\
& \pm & \sum_{n=1}^{N}\Big[\frac{1}{\lambda_n}(\delta
C_n^{\pm}-C_n^{\pm}\delta \lambda_n )F_n^{\pm}(x)
+\frac{C_n^{\pm}}{i\lambda_n}\delta \kappa_n
\dot{F}_n^{\pm}(x) \Big] \nonumber  \\
\int_{\pm \infty}^{x}\delta\sqrt{\frac{q(y)}{\omega}}\text{d}y &=&
\frac{1}{2\pi i}\int_{-\infty}^{\infty}\frac{i}{\lambda(k)}\delta
\mathcal{R}^{\pm}(k) \tilde{F}^{\pm}(x,k)\text{d}k \label{eq43a'} \\
&\pm & \sum_{n=1}^{N}\Big[\frac{1}{\lambda_n}(\delta
C_n^{\pm}-C_n^{\pm}\delta \lambda_n )\tilde{F}_n^{\pm}(x)
+\frac{C_n^{\pm}}{i\lambda_n}\delta \kappa_n
\dot{\tilde{F}}_n^{\pm}(x) \Big]\nonumber  \\
\int_{\pm \infty}^{x}\delta\sqrt{\frac{q(y)}{\omega}}\text{d}y &=&
\int_0^\infty \text{d} k \Big( \tilde{\mathcal{P}}(x,k) \delta
\phi(k) - \tilde{\mathcal{Q}}(x,k) \delta \rho(k)\Big)
\label{eq43a''} \nonumber \\ &+& \sum_{n=1}^N \left(
\tilde{P}_n(x) \delta \phi_n - \tilde{Q}_n(x) \delta \rho_n
\right). \e
\end{proposition}

\begin{proof} The expansion (\ref{eq43a}) follows from
(\ref{eq:X'}) choosing $X(x)=\delta\mathcal{Q}_\pm(x)$. The
corresponding expansion coefficients (\ref{eq:X-1}),
(\ref{eq:X-1d}) are expressed in terms of the scattering data
variations using (\ref{eq:ss-1})--(\ref{eq:ss-3}). Eq.
(\ref{eq43a'}) follows from (\ref{eq43a}) and (\ref{eq:t-F}). The
expansion (\ref{eq43a''}) follows from (\ref{eq:X-s}) with
$\tilde{X}(x)=\int_{\pm \infty}^{x}\delta
\sqrt{\frac{q(y)}{\omega}} \text{d}y $. Note that condition
$\delta\alpha=0$ ensures that the left hand side of
(\ref{eq43a''}) is independent of the choice of the lower limit of
the integration. The corresponding expansion coefficients are
evaluated using the definition of the symplectic basis
(\ref{eq59})--(\ref{eq61a}), the Wronskian relations
(\ref{eq:ss-1})--(\ref{eq:ss-3}) and (\ref{eq:P-P-t}).

\end{proof}

The expansions (\ref{eq43a})--(\ref{eq43a''}) are valid for all
variations of the potential $\delta m(x)$ preserving the value of
the integral $\alpha$. An important subclass of these variations
are due to the evolution of $m(x,t)$.

Effectively we  consider a one-parameter family of spectral
problems, allowing a dependence on the additional parameter $t$,
such that $m(x,t)$ is a Schwartz class function for all values of
$t$. The variation of the potential with respect to $t$ is given
by:
\begin{equation}\label{eq:dq-t}
\delta m(x,t) \equiv m(x,t+\delta t)-m(x,t) \simeq m_t\delta t +
\mathcal{O}((\delta t)^2).
\end{equation}
For such potentials the corresponding scattering data, e.g.
$\mathcal{R}^\pm (k,t)$, $C_n^\pm(t)$, $\kappa_n$ generically will
depend also on $t$. Keeping only the first order terms with
respect to $\delta t$ we find that the corresponding variations of
the scattering data are given by:
\begin{equation}\label{eq:d-R}
\delta \mathcal{R}^\pm (k) = \mathcal{R}_t^\pm (k) \delta t +
\mathcal{O}((\delta t)^2), \qquad \delta C_n^\pm = C_{n,t}^\pm
\delta t + \mathcal{O}((\delta t)^2).
\end{equation}
With all these explanations from Proposition \ref{pr6} one easily
proves the following

\begin{cor}\label{cor:11}
\b \frac{\omega}{\sqrt{q(x)}}\int_{\pm
\infty}^{x}(\sqrt{q(y)})_t\text{d}y & = & \frac{1}{2\pi
i}\int_{-\infty}^{\infty}\frac{i}{\lambda(k)}
\mathcal{R}_t^{\pm}(k)F^{\pm}(x,k)\text{d}k \label{eq43A}  \\
& \pm & \sum_{n=1}^{N}\Big[\frac{1}{\lambda_n}(
C_{n,t}^{\pm}-C_n^{\pm} \lambda_{n,t} )F_n^{\pm}(x)
+\frac{C_n^{\pm}}{i\lambda_n} \kappa_{n,t}
\dot{F}_n^{\pm}(x) \Big] \nonumber  \\
\int_{\pm \infty}^{x}\left( \sqrt{\frac{q(y)}{\omega}} \right)_t
\text{d}y &=& \frac{1}{2\pi i}\int_{-\infty}^{\infty}
\frac{i}{\lambda(k)} \mathcal{R}_t^{\pm}(k) \tilde{F}^{\pm}(x,k)
 \text{d}k \label{eq43A'} \\
&\pm & \sum_{n=1}^{N}\Big[\frac{1}{\lambda_n}(
C_{n,t}^{\pm}-C_n^{\pm} \lambda_{n,t} )\tilde{F}_n^{\pm}(x)
+\frac{C_n^{\pm}}{i\lambda_n}\kappa_{n,t}
\dot{\tilde{F}}_n^{\pm}(x) \Big]\nonumber  \\
\int_{\pm \infty}^{x}\left(\sqrt{\frac{q(y)}{\omega}}\right)_t
\text{d}y &=& \int_0^\infty \text{d} k \Big(
\tilde{\mathcal{P}}(x,k) \phi_t(k) - \tilde{\mathcal{Q}}(x,k)
\rho_t(k)\Big) \label{eq43A''} \nonumber \\ &+& \sum_{n=1}^N
\left( \tilde{P}_n(x) \phi_{n,t} - \tilde{Q}_n(x) \rho_{n,t}
\right). \e

\end{cor}

\begin{proposition}\label{pr12}
Let $\Omega(z)$ be a rational function such that its poles lie
outside the spectrum $\bbbr \cup \mathop{\cup}\limits_{n=1}^N \{
i\kappa_n, -i\kappa_n\}$. Then:
\begin{eqnarray}\label{eq36A}
\Omega(L_\pm) \Big(\sqrt{\frac{\omega}{q(x)}}-1\Big) &=&
\pm\frac{1}{2\pi i}\int_{-\infty}^{\infty}\frac{2k
\mathcal{R}^{\pm}(k)} {\omega \lambda(k)} \Omega(\lambda^{-1})
F^{\pm}(x,k)\text{d}k \nonumber \\ &+&
\sum_{n=1}^{N}\frac{2\kappa_n}{\omega\lambda_n}C_n^{\pm}
\Omega(\lambda_n^{-1})F_n^{\pm}(x),
\end{eqnarray}
\begin{eqnarray}\label{eq36B}
\Omega(\tilde{L}_\pm)\Big(1- \sqrt{\frac{q(x)}{\omega}}\Big) &=&
\pm\frac{1}{2\pi i}\int_{-\infty}^{\infty}\frac{2k
\mathcal{R}^{\pm}(k)} {\omega \lambda(k)}
\Omega(\lambda^{-1})\tilde{F}^{\pm}(x,k)\text{d}k \nonumber \\ &+&
\sum_{n=1}^{N}\frac{2\kappa_n}{\omega\lambda_n}C_n^{\pm}
\Omega(\lambda_n^{-1})\tilde{F}_n^{\pm}(x),
\end{eqnarray}
\begin{eqnarray}\label{eq72AA}
\Omega(L)\left( \sqrt{\frac{\omega}{q(x)}}-1 \right)
=-2\int_{0}^{\infty}k \Omega(\lambda^{-1})\mathcal{P}(x,k)
\text{d}k +2\sum_{n=1}^{N}\kappa_n \Omega(\lambda_n^{-1})P_n (x),
\end{eqnarray}
\begin{eqnarray}\label{eq72AA'}
\Omega(\tilde{L}) \left(1- \sqrt{\frac{q(x)}{\omega}}\right) =
-2\int_{0}^{\infty}k \Omega(\lambda^{-1}) \tilde{\mathcal{P}}(x,k)
\text{d}k +2\sum_{n=1}^{N}\kappa_n \Omega(\lambda_n
^{-1})\tilde{P}_n (x).
\end{eqnarray}
\end{proposition}

\begin{proof}
Eq. (\ref{eq36A}) is obtained by acting on the expansion
(\ref{eq36}) with the operator $\Omega(L_\pm)$ and using
(\ref{eq45}). The condition imposed on the poles of $\Omega(z)$
ensures that $\Omega(\lambda^{-1})$ and $\Omega(\lambda_n^{-1})$
are all finite, so the right hand side of (\ref{eq36A}) is well
defined. Eq. (\ref{eq36B}) follows from (\ref{eq36b}) and
(\ref{eq45'}). Eqs. (\ref{eq72AA}) and  (\ref{eq72AA'}) are
derived analogously using the expansions  (\ref{eq72aa}) and
(\ref{eq72aa'}) and (\ref{eq:L-P}).

\end{proof}

\begin{cor} If in Proposition \ref{pr12}  we use the operators
$\Omega(\hat{L}_\pm)$,  $\Omega(\hat{\tilde{L}}_\pm)$,
$\Omega(\hat{L})$ and  $\Omega(\hat{\tilde{L}})$ respectively,
then in the right hand sides of (\ref{eq36A})--(\ref{eq72AA'}) the
factors $\Omega(\lambda^{-1})$ and $\Omega(\lambda_n^{-1})$ will
be replaced by $\Omega(\lambda)$ and $\Omega(\lambda_n)$
respectively.

\end{cor}

\begin{proof}
Follows easily from the arguments used in Proposition \ref{pr12}
and (\ref{eq45i}) and (\ref{eq45i'}). \end{proof}

Now it is easy to describe the hierarchy of Camassa-Holm
equations. To every choice of the function $\Omega(z)$, known also
as the dispersion law we can put into correspondence the Nonlinear
Evolution Equation (NLEE):
\begin{equation}\label{eq:CH}
\frac{2}{\sqrt{q}}\int_{\pm \infty}^{x}(\sqrt{q})_t \text{d}y
+\Omega(\mathcal{L})\Big(\sqrt{\frac{\omega}{q}}-1\Big)=0,
\end{equation}
\begin{equation}\label{eq:CH-t}
\int_{\pm \infty}^{x}(\sqrt{q})_t \text{d}y
+\Omega(\tilde{\mathcal{L}})\Big(1-\sqrt{\frac{q}{\omega}}\Big)=0,
\end{equation}
where $\mathcal{L}$ (resp.  $\tilde{\mathcal{L}}$) is any of the
operators $L_+$, $L_-$ or $L$ (resp. $\tilde{L}_+$, $\tilde{L}_-$
or $\tilde{L}$).

What will be demonstrated below is that the hierarchy
(\ref{eq:CH}) (resp. (\ref{eq:CH-t})) can be generated by each of
the recursion operators $L_\pm$, $L$, (resp. $\tilde{L}_\pm$,
$\tilde{L}$) and their inverse. The fact that the expansions over
the squared solutions provide the spectral decompositions of the
recursion operators makes evident the interpretation of the ISM as
a generalized Fourier transform. Using these expansions we will
show that each CH-type equation is equivalent to a linear
evolution equation for the scattering data.

\begin{proposition}\label{pr8}
\begin{description} \item[i)]
Each of the NLEE (\ref{eq:CH}) and (\ref{eq:CH-t}) is equivalent
to the following linear evolution equations for the scattering
data: \b \mathcal{R}_t^{\pm}(k)\mp i k \Omega(\lambda^{-1})
\mathcal{R}^{\pm}(k)=0, \label{eq48}\\
C_{n,t}^{\pm}\pm \kappa_n \Omega(\lambda_n^{-1})C_n^{\pm}=0,\label{eq49} \\
\kappa_{n,t}=0.\label{eq50}\e

\item[ii)] Each of the NLEE (\ref{eq:CH}) and (\ref{eq:CH-t}) is
equivalent to the following linear evolution of the action-angle
variables: \b \phi_t(k,t)- k \Omega(\lambda^{-1}) =0, \qquad
\rho_t(k)=0, \label{eq48AA}\\
\phi_{n,t} +\kappa_n \Omega(\lambda_n^{-1})=0, \qquad
\kappa_{n,t}=0. \label{eq49}
 \e
\end{description}

\end{proposition}

\begin{proof}
i) Let us consider the NLEE (\ref{eq:CH})  fixing up
$\mathcal{L}=L_\pm$ and let us expand the left hand side of
(\ref{eq:CH}) over the squared solutions $F^\pm (x,k,t)$ using the
expansions (\ref{eq43A}) and (\ref{eq36A}). It is easy to check
that the corresponding expansion coefficients coincide with the
left hand sides of (\ref{eq48})--(\ref{eq50}). It remains to make
use of i) in Lemma \ref{lem:2}. To prove the equivalence of
(\ref{eq:CH-t}) to the linear eqs. (\ref{eq48})--(\ref{eq50}) we
fix up $\tilde{\mathcal{L}}=\tilde{L}_\pm$ and make use of the
expansions (\ref{eq43A'}), (\ref{eq36B}) and of ii) in Lemma
\ref{lem:2}.

ii) Now we choose $\tilde{\mathcal{L}}=\tilde{L}$ and expand the
left hand side of (\ref{eq:CH-t}) over the symplectic basis using
(\ref{eq48AA}), (\ref{eq49}). It remains to apply Lemma
\ref{lem:2'}.
\end{proof}

The NLEE (\ref{eq:CH}) and (\ref{eq:CH-t}) can be simplified as
follows: \b q_t+ \sqrt{q}\Big[\sqrt{q}\Omega(\mathcal{L})
\Big(\sqrt{\frac{\omega}{q}}-1\Big)\Big]_x=0, \label{eq47} \\
 \label{eq47'}
 q_t+ \Big[\Omega(\tilde{\mathcal{L}})
 \Big(1-\sqrt{\frac{q}{\omega}}\Big)\Big]_x=0,
\e

\begin{example}
With $\Omega(z)=z$ one can  easily check that \b
L_{\pm}\Big(\sqrt{\frac{\omega}{q}}-1\Big)=2u\label{Lpm}\e and
thus the equation (\ref{eq47}) becomes the Camassa-Holm equation
(\ref{eq1}). \end{example}

 The higher degree polynomials in (\ref{eq47})
produce the other members of the Camassa-Holm hierarchy. All
members of the hierarchy share the same spectral problem
(\ref{eq3}), and thus their solutions have the same
$x$-dependence. The only difference is the time-evolution of the
scattering data of the members of the hierarchy.

Now it is clear how to extend the dispersion law $\Omega(z)$ for
the case of ratio of two polynomials:

\begin{cor}\label{pr8a} Let
\b \label{eq:Ome} \Omega(z)=\frac{\Omega_2(z)}{\Omega_1(z)}.\e
where $\Omega_1(z)$ and $\Omega_2(z)$ are two polynomials. The
corresponding NLEE can be written in the form: \b
\Omega_1(L_{\pm}) \frac{2}{\sqrt{q}}\int_{\pm
\infty}^{x}(\sqrt{q})_t \text{d}y
+\Omega_2(L_{\pm})\Big(\sqrt{\frac{\omega}{q}}-1\Big)=0
\label{eq:pr8a} \e is also equivalent to (\ref{eq48}) --
(\ref{eq50}) where $\Omega (z)$ is given by (\ref{eq:Ome}).
\end{cor}

\begin{example}
$\Omega_1(z)=z$ and $\Omega_2(z)=1$, i.e. $\Omega(z)=1/z$. The
equation (\ref{eq:pr8a}) due to (\ref{eq44}) has the form

\b q_t+\partial_x(\partial_x^2-1)\sqrt{\frac{\omega}{q}}=0,
\label{eq:Fokas} \e

\n which is exactly the extended Dym equation
\cite{CH93,FOR96,OR96}. \end{example}

Of course, further generalizations are possible e.g. by
introducing another time-like variable, see \cite{Cal}.

\begin{cor}
The functional derivative $\frac{\delta f}{\delta m(x)}$ can be
expanded over the symplectic basis as follows:

\b \frac{\delta f}{\delta m(x)} =
\int_{0}^{\infty}\Big(\{f,\phi(k)\}
\mathcal{P}(x,k)-\{f,\rho(k)\}\mathcal{Q}(x,k)\Big)\text{d}k \nonumber\\
+\sum_{n=1}^{N}\Big(\{f,\phi_n\}P_n
(x)-\{f,\rho_n\}Q_n(x)\Big).\label{eq71}\e
\end{cor}

\begin{proof}
Insert $X(x)=\frac{\delta f}{\delta m(x)}$ into eq.
(\ref{eq:X-PQ}) and (\ref{eq:X-phi}). In view of (\ref{eq66a}) and
(\ref{eq66b}) and (\ref{eq31b}) all expansion coefficients turn
into the corresponding Poisson brackets.
\end{proof}

\begin{cor}
The Poisson bracket $\{f, g\}$ can be expressed as: \b \{f,g\} =
\int_{0}^{\infty}\Big(\{f,\rho(k)\}\{g,\phi(k)\}-
\{f,\phi(k)\}\{g,\rho(k)\}\Big)\text{d}k \nonumber\\
+ \sum_{n=1}^{N}\Big(\{f,\rho_n\}\{g,\phi_n\}
-\{f,\phi_n\}\{g,\rho_n\}\Big).\label{eq70}\e
\end{cor}

\begin{proof}
Multiply both sides of (\ref{eq70}) by $\frac{\delta g}{\delta
m(x)}$ and take the skew-symmetric product. Due to (\ref{eq31b})
the left hand side becomes $\{f,g\}$. The right hand side follows
from (\ref{eq:X-phi}) and (\ref{eq66a}),  (\ref{eq66b}).

\end{proof}

\begin{cor}
From (\ref{eq71}) with $f=H_1=\frac{1}{2}\int mu \text{d}x$ and
the relations $\{\phi,H_1\}=k/\lambda(k)$,
$\{\phi_n,H_1\}=-\kappa_n/\lambda_n$, see \cite{CI06} we obtain

\b u(x) =
-\int_{0}^{\infty}\frac{k}{\lambda(k)}\mathcal{P}(x,k)\text{d}k
+\sum_{n=1}^{N}\frac{\kappa_n}{\lambda_n}P_n (x).\label{eq72}\e
\end{cor}

\n Since the potential $m(x)$ has the meaning of 'momentum'
\cite{M98,K04}, (\ref{eq72}) gives expansion of
$u=(1-\partial_x)^{-1}m$ over the 'modes' $\mathcal{P}(x,k)$,
$P_n(x)$, representing the 'momentum' part of the symplectic
basis. Actually, (\ref{eq72}) is equivalent to (\ref{eq36}) due to
(\ref{Lpm}) and the fact that \b
L_{\pm}\mathcal{P}(x,k)=\lambda^{-1}(k)\mathcal{P}(x,k),\qquad
L_{\pm}P_n(x)=\lambda_n^{-1}P_n(x), \label{LonP}\e

\n see (\ref{eq59}), (\ref{eq61}), (\ref{eq45}).

\section{Hamiltonian formulation for the CH hierarhy}\label{sec:9}

Let us start from the following observation. From the identities
(\ref{Lpm}), (\ref{eq72}) we have

\b u&=&\frac{1}{2}L\Big(\sqrt{\frac{\omega}{q}}-1\Big),\label{Lu}\\
\e

\n and therefore, taking into consideration (\ref{LonP}),
(\ref{eq:LP2}), \b
\Omega(L_{\pm})\Big(\sqrt{\frac{\omega}{q}}-1\Big)=
\Omega(L)\Big(\sqrt{\frac{\omega}{q}}-1\Big). \nonumber \e Thus,
the equation (\ref{eq47}) can be written as \b m_t=\biglb
\delta(\bullet),\frac{1}{2}\Omega(L)
\Big(\sqrt{\frac{\omega}{q}}-1\Big)\bigrb,\label{eq80} \e

\n where $\delta(\bullet)$ denotes the delta function.

Due to (\ref{eq31b}) we can write this equation in Hamiltonian
form

\b m_t=\{m, H^{\Omega}\}, \label{eq81} \e

\n with Hamiltonian $H^{\Omega}$ such that

\b \frac{\delta H^{\Omega}}{\delta m}=
\frac{1}{2}\Omega(L)\Big(\sqrt{\frac{\omega}{q}}-1\Big).\label{eq82}
\e

From (\ref{eq72AA}) we have further \b
\frac{1}{2}\Omega(L)\Big(\sqrt{\frac{\omega}{q}}-1\Big)=
-\int_{0}^{\infty}k \Omega(\lambda^{-1})\mathcal{P}(x,k)\text{d}k
+\sum_{n=1}^{N}\kappa_n \Omega(\lambda_n^{-1})P_n (x)\nonumber
\\ =-\int_{0}^{\infty}k \Omega(\lambda^{-1})\frac{\delta \rho(k)}{\delta m(x)}\text{d}k
+\sum_{n=1}^{N}\kappa_n \Omega(\lambda_n^{-1}) \frac{\delta
\rho_n}{\delta m(x)}.\label{eq83} \e

Therefore, in view of (\ref{eq82}) and (\ref{eq83})

\b \frac{\delta H^{\Omega}}{\delta \rho(k)}=-k
\Omega(\lambda^{-1}),\qquad
 \frac{\partial H^{\Omega}}{\partial \rho_n}=\kappa_n \Omega(\lambda_n^{-1}),\label{eq84}
\e

\n and finally

\b H^{\Omega}=-\int_{0}^{\infty}k \Omega(\lambda^{-1})
\rho(k)\text{d}k -\frac{2}{\omega}\sum_{n=1}^{N}\int
\frac{\kappa_n ^2}{\lambda_n^2} \Omega(\lambda_n^{-1}) \text{d}
\kappa_n.\label{eq84x} \e

{\bf Example:} As an example we can point out the CH equation
($\Omega(z)=z$). The expression (\ref{eq84x}) gives

\b H^{\Omega}\equiv H_1=-\int_{0}^{\infty}\frac{k}{\lambda}
\rho(k)\text{d}k
+\omega^2\sum_{n=1}^{N}\Big(\ln\frac{1-2\kappa_n}{1+2\kappa_n}+\frac{4\kappa_n(1+4\kappa_n^2)}{(1-4\kappa_n^2)^2}\Big).\nonumber
\e

\n The last expression was obtained in a different way in
\cite{CI06}.$\Box$

Noticing that $\int_{\pm \infty}^x=\partial_x^{-1}$ we can rewrite
(\ref{eq2a}) as \begin{equation}\label{eq2aa} \frac{\delta
H_{n}[m]}{\delta m}=-\frac{1}{2}L_{\pm}\frac{\delta
H_{n-1}[m]}{\delta m}
\end{equation}
\n due to (\ref{eq44}), or as
\begin{equation}\label{Lenard}
\frac{\delta H_{n}[m]}{\delta m}=-\frac{1}{2}L\frac{\delta
H_{n-1}[m]}{\delta m}.
\end{equation}

\n  If  $H^{\Omega}\equiv H^{\Omega}_1$ is the Hamiltonian with
respect to the Poisson bracket (\ref{PB}), the other conservation
laws can be generated according to

\begin{equation}\label{eq2aaa}
\frac{\delta H^{\Omega}_{n}[m]}{\delta
m}=-\frac{1}{2}L\frac{\delta H^{\Omega}_{n-1}[m]}{\delta m}.
\end{equation}

For example

\b\label{eq2aaaa} \frac{\delta H^{\Omega}_{2}}{\delta
m}&=&-\frac{1}{2}L\frac{\delta H^{\Omega}_{1}}{\delta m}\nonumber \\
&=&-\frac{1}{2}L\Big(\-\int_{0}^{\infty}k \Omega(\lambda^{-1})
\mathcal{P}(x,k)\text{d}k +\sum_{n=1}^{N}\kappa_n
\Omega(\lambda_n^{-1})P_n (x)\Big)\nonumber \\
&=&-\int_{0}^{\infty}\frac{k}{(-2\lambda)}
\Omega(\lambda^{-1})\mathcal{P}(x,k)\text{d}k
+\sum_{n=1}^{N}\frac{\kappa_n}{(-2\lambda_n)}
\Omega(\lambda_n^{-1})P_n (x), \nonumber\e

\n or in general

\b\label{eq2a5} \frac{\delta H^{\Omega}_{j}}{\delta m}&=&
-\int_{0}^{\infty}(-2\lambda)^{1-j}k
\Omega(\lambda^{-1})\mathcal{P}(x,k)\text{d}k \nonumber
\\ &\phantom{*}& +\sum_{n=1}^{N}\kappa_n (-2\lambda_n)^{1-j}
\Omega(\lambda_n^{-1})P_n (x); \nonumber\e

\b \frac{\delta H_j^{\Omega}}{\delta \rho(k)}=-k(-2\lambda)^{1-j}
\Omega(\lambda^{-1}),\qquad
 \frac{\partial H_j^{\Omega}}{\partial \rho_n}=\kappa_n (-2\lambda_n)^{1-j}\Omega(\lambda_n^{-1}),\label{eq84a}
\e

\n for $j\ge 1$, or

\begin{equation} H_j^{\Omega}=-\int_{0}^{\infty}k
(-2\lambda)^{1-j}\Omega(\lambda^{-1}) \rho(k)\text{d}k
-\frac{2}{\omega}\sum_{n=1}^{N}\int \frac{\kappa_n
^2}{\lambda_n^2}(-2\lambda_n)^{1-j} \Omega(\lambda_n^{-1})
\text{d} \kappa_n.  \\ \label{eq85} \end{equation}

{\bf Example:} For the CH equation ($\Omega(z)=z$) the expression
(\ref{eq85}) gives\footnote{This is also the quantity given in
formula (81) of \cite{CI06}, however with a technical error in the
contribution from the continuous spectrum. The correct expression
should be \b
H_2=\omega^3\sum_{n=1}^{N}\Big(\ln\frac{1-2\kappa_n}{1+2\kappa_n}+\frac{4\kappa_n(3+32\kappa_n^2-48\kappa_n^4)}{3(1-4\kappa_n^2)^3}\Big)+
\frac{2^8\omega^3}{\pi}\int_{0}^{\infty}\frac{k^2\ln
|a(k)|}{(4k^2+1)^4} \text{d}k. \nonumber \e}

\begin{equation} H_2^{\Omega}\equiv H_2=\int_{0}^{\infty}\frac{k}{2\lambda^2}
\rho(k)\text{d}k
+\omega^3\sum_{n=1}^{N}\Big(\ln\frac{1-2\kappa_n}{1+2\kappa_n}+\frac{4\kappa_n(3+32\kappa_n^2-48\kappa_n^4)}{3(1-4\kappa_n^2)^3}\Big).
\end{equation}
$\Box$

Since $L^{-1}$ is well defined, it is possible to consider the
formal expression (\ref{eq85}) for $j\le0$.

{\bf Example:} For the CH equation (\ref{eq85}) gives ($j=0$,
$\Omega(z)=z$):
\begin{equation} H_0=2\int_{0}^{\infty}k
 \rho(k)\text{d}k+
2\omega\sum_{n=1}^{N}\Big(\ln\frac{1-2\kappa_n}{1+2\kappa_n}+\frac{4\kappa_n}{1-4\kappa_n^2}\Big).
\\ \label{eqH0} \end{equation}

\n  In terms of $q(x)$ we have \footnote{Note the difference in
comparison with the definitions of the integrals in \cite{CI06}.}
\b H_{0}\equiv \int_{-\infty}^{\infty}(\sqrt{q}-\sqrt{\omega
})^2\text{d}x. \label{H0} \e

\n Indeed, since $\delta H_{0}/\delta m=1-\sqrt{\omega/q}$,
$\delta H_{1}/\delta m = u$, $H_0$ and $H_1$ are related through
(\ref{eq2aa}), see (\ref{Lpm}).$\Box$

{\bf Example:} With (\ref{eq2aa}), (\ref{eq44}) one can check that
for the CH equation
\begin{equation}
H_{-1}=\frac{1}{2}\int_{-\infty}^{\infty}\Big[\Big(\sqrt[4]{\frac{\omega}{q}}-\sqrt[4]{\frac{q}{\omega}}\Big)^2+\frac{\sqrt{\omega}q_x^2}{4q^{5/2}}
\Big]\text{d}x.\label{eq90}\end{equation}

\n On the other hand, for $j=-1$ and $\Omega(z)=z$, (\ref{eq85})
gives

\begin{equation} H_{-1}=-4\int_{0}^{\infty}k
\lambda \rho(k)\text{d}k+2\sum_{n=1}^{N}
\Big(\ln\frac{1+2\kappa_n}{1-2\kappa_n}-4\kappa_n \Big).
\label{eq91}
\end{equation} $\Box$

\begin{cor}
Obviously,
$H_{1}^{\Omega(z)=1/z}=\frac{1}{4}H_{-1}^{\Omega(z)=z}$, therefore
the expressions on the right hand side of (\ref{eq90}),
(\ref{eq91}) give the Hamiltonian of (\ref{eq:Fokas}), (up to a
constant factor) with respect to the Poisson bracket
(\ref{PB}).
\end{cor}

Here we notice that for the CH equation ($\Omega(z)=z$) the
integral $\alpha$, (\ref{eqi8}), (\ref{eqi24})
\begin{equation}\label{eqi88}
\alpha \equiv
\int_{-\infty}^{\infty}\Big(\sqrt{\frac{q}{\omega}}-1\Big)\text{d}x=
\sum_{n=1}^{N}\ln\Big(\frac{1+2\kappa_n}{1-2\kappa_n}\Big)^2+
 \int _{0}^{\infty}\frac{\lambda}{k}\rho(k)\text{d}k
\end{equation}

\n is not of the form (\ref{eq85}).  However, the hierarchy is
generated by $a(k)$ and $\alpha = \lim_{k\to\infty} \frac{d}{dk}
\ln a(k)$.  Thus, $\alpha$ from (\ref{eqi88}) does not give rise
to a separate hierarchy $\{\alpha_n\}$ of conservation laws.
Another way of seeing this is to assume the contrary, that there
exists an independent hierarchy, such that

\b \frac{\delta \alpha_n}{\delta m}= L^{n}\frac{\delta
\alpha}{\delta m}, \qquad n=\pm1, \pm2, \ldots.\nonumber \e

\n However, one can check that $L\frac{\delta \alpha}{\delta m}$
and $\frac{\delta \alpha}{\delta m}$ are related as follows:

\b L\frac{\delta \alpha}{\delta m}= -4\omega \frac{\delta
\alpha}{\delta m}-2\frac{\delta H_0}{\delta m},\nonumber \e

\n and therefore no other independent integrals arise.

{\bf Example:} The following integrals are related to $\alpha$ and
an integral from (\ref{eq85}):

\b I_0&\equiv &\int_{-\infty}^{\infty}m\text{d}x= H_0+
2\omega \alpha \nonumber \\
&=& -\int _{0}^{\infty}\frac{1}{2k}\rho(k)\text{d}k +2\omega
\sum_{n=1}^{N}\Big(\ln\frac{1+2\kappa_n}{1-2\kappa_n}+\frac{4\kappa_n}{1-4\kappa_n^2}\Big),
\nonumber \\
I_{-1}&\equiv&\frac{\sqrt{\omega}}{2}\int_{-\infty}^{\infty}\Big(\frac{1}{\sqrt{q}}-\frac{1}{\sqrt{\omega}}+\frac{q_x^2}{4q^{5/2}}\Big)\text{d}x=
H_{-1}-\frac{1}{2}\alpha, \nonumber \e

\n etc. $\Box$

The equation (\ref{Lenard}) can be written as
\begin{equation}\label{eq2ab}
 \frac{\delta H_{1}^{\Omega}}{\delta m}=(-2L^{-1})^{n-1}\frac{\delta H_{n}^{\Omega}}{\delta m}
 \end{equation}
\n Therefore, defining a Poisson bracket \b
\{A,B\}_{(n)}\equiv\Bigglb \frac{\delta A}{\delta
m},(-2L^{-1})^{n-1}\frac{\delta B}{\delta m}\Biggrb, \label{PBn}
\e

\n we can write the equation (\ref{eq47}) as

\b m_t=\{m,H_n^{\Omega}\}_{(n)}.\label{eq96} \e The following
statement gives the canonical variables for (\ref{eq47}) with
respect to (\ref{PBn}).
\begin{proposition}\label{pr13} Let us define
\b \rho(k)_{(n)}&\equiv&\rho(k), \qquad
\phi (k)_{(n)}\equiv(-2\lambda)^{1-n}\phi(k) \label{eq98} \\
\rho_{l,(n)}&\equiv&\rho_l, \qquad \phi_{l,(n)}\equiv
(-2\lambda_l)^{1-n}\phi_l. \label{eq99} \e These variables satisfy
the following canonical relations with respect to the bracket
(\ref{PBn}): \b \{\rho(k_1)_{(n)},\phi
(k_2)_{(n)}\}_{(n)}&=&\delta(k_1-k_2), \nonumber \\
\{\rho(k_1)_{(n)},\rho
(k_2)_{(n)}\}_{(n)}&=&\{\phi(k_1)_{(n)},\phi
(k_2)_{(n)}\}_{(n)}=0,
\qquad k_{1,2}>0 \nonumber \\
\{\rho_{l,(n)},\phi_{p,(n)}\}_{(n)}&=&\delta_{lp}, \nonumber  \\
\{\rho_{l,(n)},\rho_{p,(n)}\}_{(n)}&=&\{\phi_{l,(n)},\phi_{p,(n)}\}_{(n)}=0.\nonumber
\e

\end{proposition}

\begin{proof}  One can perform the verification making use of
(\ref{eq:L-P}), (\ref{eq:L-Pn}). For the quantities on the
continuous spectrum the computations are particularly simple, e.g.
\b \{\rho(k_1)_{(n)},\phi
(k_2)_{(n)}\}_{(n)}&=&(-2\lambda(k_2))^{1-n}\Bigglb
\frac{\delta\rho(k_1)}{\delta m},(-2L^{-1})^{n-1}
\frac{\delta \phi (k_2)}{\delta m}\Biggrb\nonumber \\
&=&(-2\lambda(k_2))^{1-n}\Bigglb
\mathcal{P}(k_1),(-2L^{-1})^{n-1}\mathcal{Q}(k_2) \Biggrb
\nonumber\\&=& \Bigglb \mathcal{P}(k_1), \mathcal{Q}(k_2)\Biggrb
=\delta(k_1-k_2). \nonumber \e

\n For the quantities on the discrete spectrum one can notice
first that \b \frac{\delta \phi_{l,(n)}}{\delta
m}&=&\frac{\delta(-2\lambda_l)^{1-n}}{\delta m}\phi_l
+(-2\lambda_l)\frac{\delta \phi_l}{\delta m}\nonumber
\\ &=&(-2\lambda_l)^{1-n}(n-1)\lambda_l\phi_lP_l+(-2\lambda_l)^{1-n}Q_l.\label{eq100} \e

Then \b & \phantom{*}& \{\rho_{l,(n)},\phi_{p,(n)}\}_{(n)}=\nonumber \\
&=& \Bigglb \frac{\delta\rho_l}{\delta
m},(-2L^{-1})^{n-1}\frac{\delta
\phi_{p,(n)}}{\delta m}\Biggrb \nonumber \\
&=&\bigglb P_l,(-2L^{-1})^{n-1}(-2\lambda_p)^{1-n}
\Big((n-1)\lambda_p\phi_pP_p+Q_p\Big)\biggrb \nonumber \\&=&
(n-1)\lambda_p\phi_p\bigglb P_l,P_p\biggrb + \bigglb
P_l,Q_p\biggrb=\delta_{lp}, \nonumber \e \b
\{\rho_{l,(n)},\rho_{p,(n)}\}_{(n)}&=& \Bigglb
\frac{\delta\rho_l}{\delta m},(-2L^{-1})^{n-1}\frac{\delta
\rho_{p}}{\delta m}\Biggrb \nonumber \\
&=&(-2\lambda_p)^{n-1}\Bigglb P_l,P_p\Biggrb =0,\nonumber \e
\b& \phantom{*}&\{\phi_{l,(n)},\phi_{p,(n)}\}_{(n)}= \nonumber \\
&=&(-2\lambda_l)^{1-n}\Bigglb (n-1)\lambda_l\phi_l P_l+Q_l,
(n-1)\lambda_p\phi_pP_p+Q_p\Biggrb  \nonumber \\
&=&(-2\lambda_l)^{1-n}\Big((n-1)\lambda_l\phi_l \bigglb
P_l,Q_p\biggrb +(n-1)\lambda_p\phi_p\bigglb Q_l,P_p\biggrb
\Big)\nonumber\\
&=&(-2\lambda_l)^{1-n}(n-1)( \lambda_l\phi_l-\lambda_p\phi_p)
\delta_{lp}=0.\nonumber \e

\end{proof}

\section{Inverse scattering transform}\label{sec:10}

Inverse scattering method for the hierarchy (\ref{eq47}) is the
same as the one for the CH equation \cite{CGI}. The only
difference is the time-dependence of the scattering data
(\ref{eq48}) -- (\ref{eq50}). For example, the inverse scattering
is simplified in the important case of the so-called
reflectionless potentials, when the scattering data is confined to
the case $\mathcal{R}^{\pm}(k)=0$ for all real $k$. This class of
potentials corresponds to the $N$-soliton solutions of the CH
hierarchy. In this case $b(k)=0$ and $|a(k)|=1$ and $i\dot{a}_p$
is real:
\begin{equation}
i\dot{a}_p = \frac{1}{2\kappa_p}e^{\alpha\kappa_p}\prod _{n\neq
p}\frac{\kappa_p-\kappa_n}{\kappa_p+\kappa_n},\quad\text{where}\quad
\alpha = \sum
_{n=1}^{N}\ln\Big(\frac{1+2\kappa_n}{1-2\kappa_n}\Big)^2.\nonumber
\end{equation}
Thus, $i\dot{a}_p$ has the same sign as $b_n$, and therefore $
C_n^+\equiv b_n/(i\dot{a}_p) >0. $ The time evolution of $C_n^+$
is (\ref{eq49}) \b C_n^+(t)=C_n^+(0)\exp\Big(-\kappa_n
\Omega(\lambda_n^{-1})t \Big).\e

The $N$-soliton solution is \cite{CGI}
\begin{equation}
u(x,t)=\frac{\omega}{2}\int_0^{\infty} \exp\Big(-|x-g(\xi,t)|\Big)
\xi^{-2}g_{\xi}^{-1}(\xi,t)\text{d}\xi-\omega,
\label{u}\end{equation} where $g(\xi,t)$ can be expressed through
the scattering data as
\begin{eqnarray} g(\xi,t)&\equiv&\ln
\int_0^{\xi}\Big(1-\sum_{n,p}\frac{C_n^+(t)
\underline{\xi}^{-2\kappa_n}}{\kappa_n+1/2}A^{-1}_{np}
[\underline{\xi},t]\Big)^{-2}\text{d} \underline{\xi},\label{g}\e
with \b  A_{pn}[\xi,t]&\equiv&
\delta_{pn}+\frac{C_n^+(t)\xi^{-2\kappa_n}}{\kappa_p+\kappa_n}.\nonumber
\end{eqnarray}

For the peakon solutions ($\omega=0$) the dependence on the
scattering data is also known \cite{BBS98,BBS99}.

The CH multi-soliton solutions also appear in several works
\cite{CZ01,C01,J03,Li04,Li05,PI, PII, PIII,Ma05}. The Darboux
transform for the CH equation is obtained in \cite{S98}. The
construction of multi-soliton and multi-positon solutions using
the Darboux/B{\"a}cklund transform is presented in \cite{H99,PLA}.

\section{Conclusions} \label{sec:11}
In this paper the Inverse Scattering Transform for the CH
hierarchy is interpreted as a Generalized Fourier Transform. The
generalized exponents are the squares of the eigenfunctions of the
associated spectral problem. Apparently the CH hierarchy is well
defined only if $q(x,0)\equiv m(x,0)+\omega>0$. The only exception
is the CH equation itself. The situation for CH when the condition
$q(x,0)>0$ on the initial data does not hold is more complicated
and requires separate analysis \cite{K05,B04,C01,CE98,BC07}. Throughout
this work the solutions $u(x,t)$ are confined to be functions in
the Schwartz class, $\omega>0$. The inverse scattering is outlined
in detail in \cite{CGI}.

The spectral problem (\ref{eq3}) is gauge equivalent to a
standard Sturm-Liouville problem, well known from the KdV
hierarchy \b -\phi_{yy}+U(y)\phi&=&\mu \phi, \qquad
\mu = -\frac{1}{4\omega}-\lambda, \nonumber \\
\phi(y)&=&q^{1/4}\Psi, \qquad \frac{dy}{dx}=\sqrt{q},
\label{chvar} \\
U(y)&=&\frac{1}{4q(y)}+\frac{q_{yy}(y)}{4q(y)}-\frac{3q_y^2(y)}{16q^2(y)}-\frac{1}{4\omega}.
\label{EP} \e

\noindent
Note that (\ref{chvar}) leads to two possible expressions for the
change of the variables in the Liouville transformation: \b y&=&
\sqrt{\omega}x+\int_{-\infty}^x (\sqrt{q(x')}-\sqrt{\omega}){\text
d}x'+ \text{const}, \label{y1} \\ y&=&
\sqrt{\omega}x+\int_{\infty}^x (\sqrt{q(x')}-\sqrt{\omega}){\text
d}x'+ \text{const}. \label{y2} \e

\n These two possibilities, (\ref{y1}), (\ref{y2}) are only
consistent iff \b \int_{-\infty}^{\infty}
(\sqrt{q(x)}-\sqrt{\omega}){\text d}x=\text{constant}, \nonumber \e
which is always the case, since the integral under question is (up
to a multiplier) the Casimir function $\alpha$ (\ref{eqi8}); see
some details in \cite{L02}.

The matching of the CH hierarchy to KdV hierarchy requires solving
the Ermakov-Pinney equation (\ref{EP}) \cite{C01,CL}, which is not
straightforward. One can eventually obtain a solution in
parametric form \cite{J03,Li04}, see also \cite{PI,PII,PIII,Ma05}.
The analytic properties of the eigenfunctions and especially their
asymptotics for $k\to\infty $ in these two cases are substantially
different \cite{CGI}, e.g. compare (\ref{eq21aa}), (\ref{eq22})
with the well known results for the standard Stourm Liouville
problem $ e^{ikx} (1 + ... )$. This influences also the dispersion
relation for the transmission coefficient. Thus the matching of
the IST for these two cases is not automatic. This alternative
approach relies on several implicit equations and is considerably
less transparent than the approach adopted here.

We have also excluded the possibility of 'creation' or 'death' of
solitons, i.e. an appearance of a new discrete eigenvalue as a
result of an infinitesimal change $\delta m$ in $m$. For the KdV
equation this problem is addressed e.g. in \cite{E81}.

The behavior of the scattering data at $k=0$ is also an important
question. In our analysis we implicitly assumed that the Wronskian
\b W(f^+(x,k),f^-(x,k))|_{x=0} \ne  0. \nonumber \e Then  $a(k)$
at $k=0$ has a singularity of type $k^{-1}$, cf. (\ref{eq10a}).
However it is possible that $W(f^+(x,k),f^-(x,k))|_{x=0}= 0$, and
then $a(k)$ is not singular at $k=0$. To investigate the behavior
of the quantities at $k=0$ one can proceed as in \cite{E81}. There
is a basis in the space of eigenfunctions of the spectral problem,
which can be chosen as $f_1(x)=f^+(x,0)$ and
$f_2(x)=-i\dot{f}^+(x,0)$. The asymptotics are

\b f_1(x)\rightarrow 1, \quad x\rightarrow \infty;\qquad
f_1(x)\rightarrow Ax+B, \quad x\rightarrow -\infty;\nonumber \\
f_2(x)\rightarrow x, \quad x\rightarrow \infty;\qquad
f_2(x)\rightarrow Cx+D, \quad x\rightarrow -\infty,\nonumber \e

\n where $A$, $B$, $C$, $D$ do not depend on $x$, e.g.

\b A&=&\frac{1}{4\omega} \int _{-\infty}^{\infty}
m(x)f_1(x)\text{d}x, \nonumber \\
B&=&1-\frac{1}{4\omega} \int _{-\infty}^{\infty} x
m(x)f_1(x)\text{d}x \nonumber \\  C&=&1+\frac{1}{4\omega} \int
_{-\infty}^{\infty} m(x)f_2(x)\text{d}x.\e

The Wronskian has the same value at $-\infty$ and $+\infty$,
therefore \b BC-AD=1. \label{Wronskian1}\e The behavior of the
scattering data is

\b a(k)&=&\frac{A}{2ik}+\frac{B+C}{2}+o(1), \label{a} \\
b(k)&=&-\frac{A}{2ik}+\frac{C-B}{2}+o(1). \label{b} \e

Note that $A$ is an integral of motion (cf.
\cite{APP87,APP88,BK89}) as well as $B+C$ since $a(k)$ does not
depend on $t$. If $A=0$ , the singularity disappears. Then, from
(\ref{Wronskian1}) it follows that $BC=1$ (and $B$ and $C$ are
integrals of motion by themselves in this case -- since both $B+C$
and $BC$ are), i.e. if $C=B^{-1}=\sigma e^{\beta}$, $\sigma=\pm1$,
then

\b a(0)=\sigma \cosh \beta,\qquad b(0)=\sigma \sinh \beta, \qquad
\mathcal{R}(0)=\tanh \beta. \nonumber \e

This situation, although exceptional is the one in which the
purely solitonic case is allowed: when $\beta=0$,
$\mathcal{R}(0)=0$ (if $A\ne 0$, $\mathcal{R}(0)=-1$, see
(\ref{a}), (\ref{b})).

The Poisson bracket (\ref{PBa}) is defined through variations with
respect to $m$. We established that among these variations there
are some that do not vanish in the asymptotic limit
$|x|\rightarrow\infty$, such as for example the variations of the
scattering data. This fact (which is not related to the smoothness
or rate of decay of $m(x)$) is related to the presence (in
general) of poles of $a(k)$ and $b(k)$ at $k=0$. More careful
analysis \cite{APP87,APP88,FT85} leads to a modification in the
definition of the Poisson bracket by additional terms, when the
behavior of variations like $\delta a(k)/\delta m$, $\delta
b(k)/\delta m$ is considered at $k=0$.

\section*{Acknowledgements}

A.C. acknowledges funding from the Science Foundation Ireland,
Grant 04 BRG/M0042. V.S.G. acknowledges funding from the Bulgarian
National Science Foundation, Grant 1410, R.I.I. acknowledges
funding from the Irish Research Council for Science, Engineering
and Technology. The authors are grateful to both referees for very
helpful suggestions.

\end{document}